\documentclass[fleqn,10pt,table]{wlscirep}
\usepackage[utf8]{inputenc}
\usepackage[T1]{fontenc}
\usepackage{bm}
\usepackage{amssymb}
\usepackage{amsmath}
\usepackage{graphbox}
\usepackage{xfrac}
\usepackage{xcolor}
\AtBeginEnvironment{tabular}{\small}

\definecolor{darkgreen}{rgb}{0,0.80,.0}

\newcommand\mat[1]{{\color{blue}#1}}

\title{Quantum mixtures of ultracold gases of neutral atoms}

\author[1,2]{Cosetta Baroni}
\author[1]{Giacomo Lamporesi}
\author[3,4]{Matteo Zaccanti}

\affil[1]{Pitaevskii BEC Center, CNR-INO and Dipartimento di Fisica, Universit\`a di Trento, 38123 Trento, Italy}
\affil[2]{Institute for Quantum Optics and Quantum Information (IQOQI), Austrian Academy of Sciences, 6020 Innsbruck, 
Austria}
\affil[3]{Istituto Nazionale di Ottica del Consiglio Nazionale delle Ricerche (CNR-INO), 50019 Sesto Fiorentino, Italy}
\affil[4]{European Laboratory for Non-Linear Spectroscopy (LENS), Universit\`{a} di Firenze, 50019 Sesto Fiorentino, Italy}

\affil[*]{e-mail: cosetta.baroni@ino.cnr.it, giacomo.lamporesi@ino.cnr.it,  zaccanti@lens.unifi.it}

\begin{abstract}
{ 
After decades of improvements in cooling techniques of several atomic species and in finding methods for the achievement of stable quantum mixtures, the field is now ready for an extensive use of such a versatile experimental platform for the investigation of a variety of physical problems. Among them, relevant examples are the dynamics of impurities in a quantum gas, the miscibility condition of different gases, the study of exotic topological structures, the interplay between magnetism and superfluidity, the formation of artificial molecules, or new few-body states.
We illustrate the differences among possible quantum mixtures, be they homonuclear spin mixtures or heteronuclear ones, and show how they can be exploited to investigate a plethora of topics from the few-body to the many-body regime. 
In particular, we discuss quantum mixtures of ultracold gases under three different perspectives: systems made of a few atoms of different kinds, single impurities immersed in a host quantum gas, and quantum mixtures of two interacting gases.
We restrict the discussion to single harmonic or flat traps, predominantly in a three-dimensional configuration.
A selection of results on recent experiments and possible interesting future directions are given.  } 
\end{abstract}

\begin{document}

\flushbottom
\maketitle

\thispagestyle{empty}

\section*{Introduction}
One hundred years ago Satyendranath Bose \cite{Bose1924pgu} and Albert Einstein \cite{Einstein1924qde} theoretically predicted that a gas of (then named) bosons can condense in a single quantum state, producing a macroscopic quantum system with long-range coherence and superfluid properties \cite{London1938tlp}: the so-called Bose-Einstein condensate (BEC). Shortly after, Enrico Fermi \cite{Fermi1926sqd} and Paul Dirac \cite{Dirac1926ott} derived a statistical formalism to describe the behaviour of "fermions", particles subjected to the exclusion principle formulated shortly before by Wolfgang Pauli \cite{Pauli1925udz}. With these two quantum statistics, we are able to theoretically understand the quantum properties of atomic systems, which can be produced and studied in ultracold atoms laboratories with exquisite control.

Thanks to the rapid advancements in the manipulation techniques of atomic internal and external degrees of freedom with laser cooling \cite{Schreck2021lcf}, the first BECs were experimentally produced in 1995 at JILA in the group of C. Wieman and E. Cornell \cite{Anderson1995oob}, and at MIT in the group of W. Ketterle \cite{Davis1995bec}.
In 1999 also the first degenerate Fermi gas was produced at JILA in the group of D. Jin, using similar techniques \cite{DeMarco1999oof}. Since then, gases of many atomic species have been successfully cooled down to quantum degeneracy, allowing for direct observation of several fundamental quantum phenomena \cite{Ketterle1999mpa,Ketterle2008mpa,Bloch2008mbp,Zwerger2012tbb}.

The improvement in the level of atomic control, also thanks to the observations of Feshbach resonances~\cite{Chin2010fri, Stan2004oof, Inouye2004ooh}, which allow the tuning of the interparticle interactions, soon led to the idea of combining atoms of different kinds in such extreme conditions and realize ultracold quantum mixtures. Such systems opened unprecedented possibilities of investigating paradigmatic physics phenomena in diverse research directions, encompassing few- and many-body fundamental problems, such as the realization of ultracold gases of 
heteronuclear polar molecules \cite{Carr2009cau}, the behaviour of impurities in an ultracold gas \cite{Schirotzek2009oof, Spethmann2012dos}, the exploration of the BCS-BEC crossover in fermionic superfluid mixtures \cite{Zwerger2012tbb}, or the miscible-immiscible phase transition in bosonic superfluid mixtures \cite{Timmermans1998pso,Ao1998bbe,Trippenbach2000sob}, just to mention a few relevant examples. 

This review, targeted at a non-expert reader, aims to introduce the most representative phenomena that can (or could potentially) be explored with such systems, and to provide a brief but comprehensive overview on past and ongoing research on quantum mixtures of ultracold atoms.
Owing to the impressive development of this research area and its ramifications into several distinct sub-fields, the selection of 'relevant' work, and the identification of the 'main concepts' regarding quantum mixtures is unavoidably subjective, and partly biased by the specific expertise of the authors \cite{Grimm2023fqm,Lamporesi2023tcs,Zaccanti2023mif}.
In particular, our work mainly focuses on three-dimensional, bulk mixtures of ultracold gases of neutral atoms, although references to research not strictly falling in this category are provided throughout the text.
Specifically, we refer the interested reader to extensive reviews, already available in the literature, for what concerns the stand-alone but connected areas of 
quantum mixtures in optical lattices \cite{Lewenstein2007uag, Bloch2008mbp,Gross2017qsw,Schäfer2020},
ultracold polar molecules \cite{Carr2009cau, Quéméner2012,Langen2023quantum}, systems in reduced dimensionality \cite{Sowinski2019odm, Mistakidis2023fbb}, Rydberg atoms \cite{Browaeys2020mbp},
and "hybrid" systems \cite{Tomza2019RMP,Lous2022},
only briefly highlighted in the present article.

Our work is organized as follows: after classifying the various types of quantum mixtures into a few main categories, we provide a general overview of the rich phenomenology associated with such many-body systems. We then discuss in more detail quantum mixtures from the three viewpoints of few-particle systems, impurity problems, and many-body physics, highlighting, in the final part,  future perspectives and possible new directions of this wide research field.

\section*{Key points}
\begin{itemize}
    \item Mixtures of atoms can be realised either using different 'spin' states of the same atomic species (\textbf{homonuclear mixtures}) or different atomic species (\textbf{heteronuclear mixtures}). The main differences concern the different level of experimental complexity, the possibility to independently manipulate the two components, the quantum statistics and mass ratio of the constituents, and the interconversion between the two components of the mixture.
    \item The combination of ultralow collision energies, quantum statistics of the scattering particles, and their mass asymmetry promotes the appearance of a variety of novel \textbf{few-particle} cluster states, exhibiting significantly different types of spatial correlations and spectral properties. 
    \item When one atom (\textbf{impurity}) is embedded in a bath of different particles with which it interacts, it dresses itself with excitations of the bath, and can be described in terms of a quasi-particle, that is an object that behaves as a free particle but with renormalized properties, such as energy and mass.
    \item The behaviour of two \textbf{bosonic gases} is mainly given by the combination of its two-body interaction strengths, with interesting features for small values of intercomponent interactions. 
    For repulsive interspecies interaction, the gases can either mix, self-organize in spatially separated domains, whereas for increasing attractive ones they can either form stable solitons or quantum droplets, or collapse.  
    \item The physics of two distinguishable \textbf{fermionic gases} is dominated by the crossover from a superfluid of tightly bound molecules to a superfluid of loosely bounded pairs while passing from strongly repulsive to strongly attractive intercomponent interactions.    
\end{itemize}

\section*{From homonuclear to heteronuclear mixtures}
At least two distinguishable kinds of particles are needed to realize an atomic mixture.
Depending on their constituents, it is possible to engineer ultracold mixtures with very different features. Two big categories are represented by {\textit{homonuclear (spin) mixtures}}  and {\textit{heteronuclear mixtures}}, as illustrated in Fig.\ref{Fig:cartoon}a. In the Supplementary Information, we provide a not exhaustive list of ultracold mixtures investigated in various experiment.
For the sake of simplicity, here we will restrict the following discussion to the case of only two components, unless differently explicitly specified.

 \begin{figure}[t!]
        \centering
        \includegraphics[width= 0.8
        \linewidth]{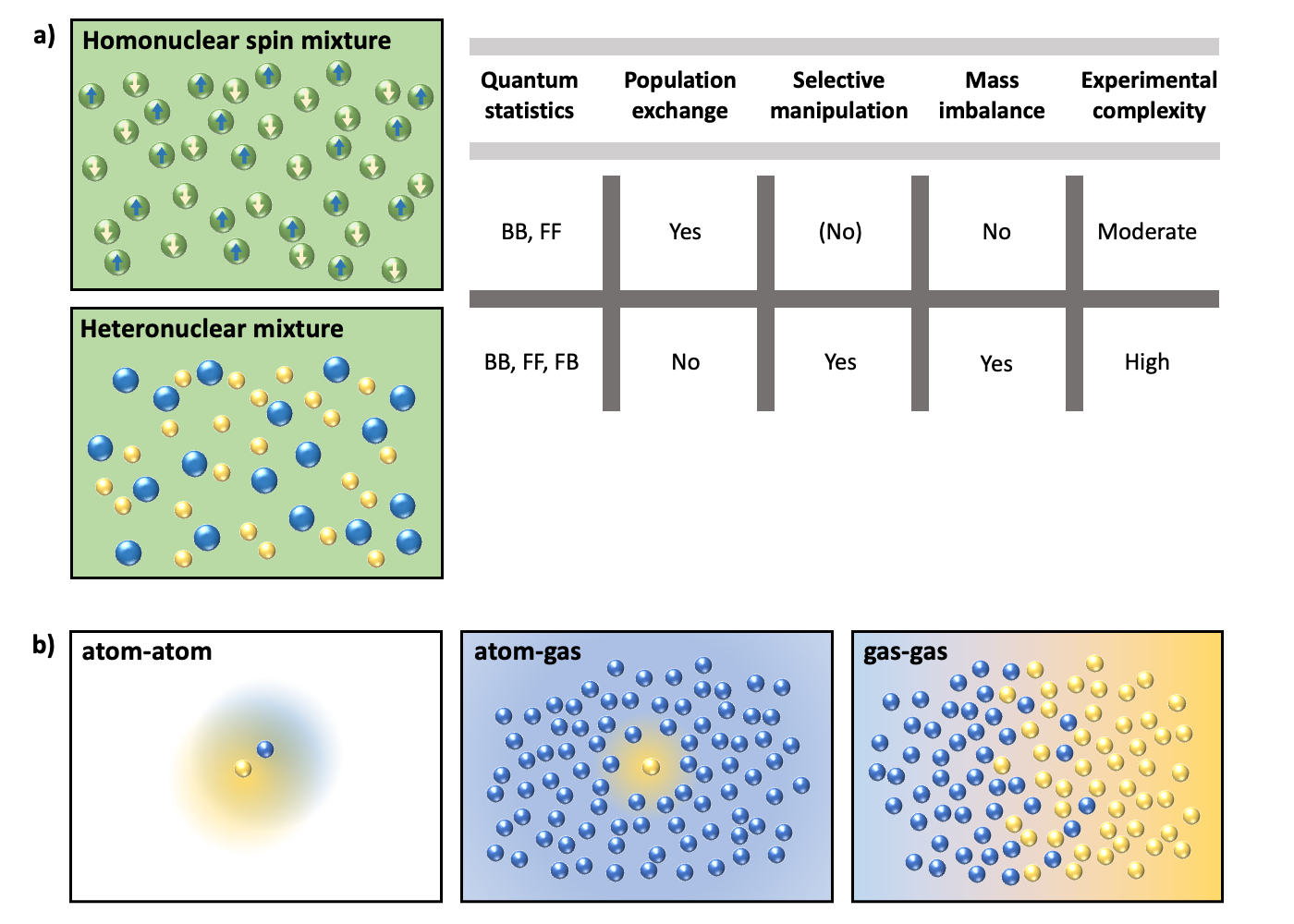}\\
        \caption{{\textbf{Schematic representation of quantum mixtures with different constituents. }}
a) Mixture made of atoms of a given species in two different spin states (up and down arrows) in the top panel, and mixture made of atoms of different atomic species in the bottom panel. Blue and yellow indicate the two different components. Different atom sizes qualitatively indicate different atomic species. Here we illustrate the case of miscible gases, with a uniform average density (green background). The table summarizes the differences between the two types of mixtures in terms of the possibility to have different quantum statistics, state interconversion, independent manipulation, mass imbalance, and the degree of experimental complexity to produce them. 
b) Schematic representation of three different fields of study, where the relative population changes. The first focuses on few-body (atom-atom) interactions, the second on isolated impurities immersed in a gas (atom-gas), and the third on many-body problems describing the interaction between two gasses (gas-gas). 
Blue and yellow identify the two different components. The background color indicates the local differential density of the two components.}
        \label{Fig:cartoon}
    \end{figure}
    
\subsection*{Homonuclear spin mixtures}
Spin mixtures are homonuclear systems in which all the atoms belong to the same isotope of a given element, but can occupy two different internal 'spin' states, depending on the manipulation of their internal degrees of freedom during the preparation and on their dynamical evolution driven by the experimental configuration. 
The first spin mixture of condensed bosons 
was realized at JILA with rubidium atoms occupying two different magnetically trappable hyperfine states\cite{Myatt1997pot}, shortly after the first BEC. 
A few years later, mixing together fermions in different spin states successfully led to the achievement of a degenerate Fermi gas without the need for sympathetic cooling techniques~\cite{DeMarco1999oof, Onofrio2016poo}. 

Dealing with a single atomic species, all particles 
in a spin mixture possess the same mass and quantum statistics. The main characteristic of such systems is that each atom can be prepared in a superposition of two spin states, a feature that is referred to as \textit{interconversion}, allowing for the creation of an effective two-level system. To increase the stability of such a two-level system, experimental tricks are necessary in order to avoid the decay from the desired to other unwanted internal spin states due to exothermal processes such as spin exchange or dipolar relaxation. Experimental techniques to prevent such a decay rely on the exploitation of the quadratic Zeeman effect and light shifts \cite{Gerbier2006rco}.
Furthermore, it is possible to introduce a coherent coupling between the two states of interest and realize systems with intriguing spin dynamics, which permit the quantum simulation of solid-state \cite{Zibold2010cba,Nicklas2015oos,Cominotti2023fia,Mancini2015ooc,Stuhl2015ves, Bouhiron2024roa} and high-energy \cite{Son2002dwo,Gallemi2019dot,Frolian2022ra1} physics phenomena,  and analogue gravity \cite{Weinfurtner2007ast}.

The energy difference between the two states of interest is in the MHz or GHz range, as the two states typically belong to the same or to different hyperfine manifolds.
This property allows the experimental introduction of coherent coupling between the two states using single-photon radio frequency radiation. Given the long wavelength with respect to the extension of the atomic samples, resulting in an uniform field along the sample, this technique makes it possible to address all the atoms at the same time. Moreover, the momentum transferred to the atom in this process is negligible.
Another possibility is the use of two-photon optical Raman transitions \cite{CohenTannoudji2011}, which can provide, contrary to the previous method, a selective local coupling if tightly focused laser beams are used. In addition, if the two beams co-propagate, the associated momentum transfer is negligible and the coupling is equivalent to a single-photon radio frequency transitions. In the case of counter-propagating beams, instead a large momentum transfer is imparted to the atoms, and internal and external degrees of freedom become intertwined, realizing so-called spin-orbit coupled systems \cite{Lin2011soc,Recati2022ccm}.

Gases of atoms occupying all the states of a hyperfine level realize a 
system that shows spinor features \cite{StamperKurn2013sbg}. 
Spinor gases have been experimentally realized in optical traps using sodium \cite{Stenger1998sdi,JimenezGarcia2019sfa}
or rubidium atoms \cite{Barrett2001aof,Chang2004oos}, revealing different ground state configurations depending on the intra- and intercomponent interactions. 
Fermionic spinor gases with more than two components have been studied with ytterbium atoms, realizing strongly correlated systems with tunable SU(N) symmetry \cite{Pagano2014aod}.

\subsection*{Heteronuclear mixtures}
Atoms of different species, or different isotopes of a given element, can be cooled down and confined together, realizing an ultracold heteronuclear mixture. 
Each component of the mixture is characterized by a well-defined atomic mass and statistics, and there is no constraint on the possibility to realize any combination of them.
All kinds of statistical mixtures have been realized using different combinations of atomic species, Bose-Bose (BB), Fermi-Fermi (FF), and even Fermi-Bose (FB). While the first two can also be obtained in spin mixtures, the implementation of the latter requires necessarily 
different atomic species, or at least different isotopes.

The first heteronuclear mixtures of  quantum-degenerate gases were realized in 2001 in Florence with a BB combination of $^{41}$K and $^{87}$Rb atoms \cite{Modugno2001bec}. In the same year, in Paris, BECs of $^{7}$Li were immersed in a single spin state Fermi sea of $^{6}$Li, forming a quantum-degenerate FB mixture\cite{Schreck2001qbe,Truscott2001oof}. Only in 2008, the first quantum-degenerate two-species FF mixtures were obtained with $^6$Li--$^{40}$K in Munich~\cite{Taglieber2008qdt} and Innsbruck~\cite{Wille2008eau}. 

The masses of the constituents can be chosen with ratios that range from approximately one, for instance in the case of isotopic mixtures,
to more than 20 for\mat{, e.g., } Li--Cs\cite{Tung2013umo} and Li--Yb\cite{Green2020fri}. Systems with a large mass imbalance are ideally suited, for example, to explore Efimov physics thanks to the  much smaller ratio between consecutive resonances compared to homonuclear systems \cite{Naidon2017epa}, or to study FF mixtures with unmatched Fermi surfaces and asymmetric dispersion relations, leading to exotic types of superfluidity \cite{Chevy2010ucp, Baarsma2010pam}. A priori, different species 
interact in different ways with optical potentials, depending on their own atomic level structure and on the  laser wavelength \cite{Grimm2000odt}. 
On the one hand, this represents an experimental complication when one aims at manipulating both species in the same way, but it can also provide a crucial extra degree of freedom for a species-dependent control \cite{LeBlanc2007sso,Catani2009eei,Lamporesi2010sim,Wilson2021qdm}.
The schematics in Fig.~\ref{Fig:cartoon}a summarizes the main differences between homo- and heteronuclear mixtures.

Only a few experiments have been reported on mixtures made  of three atomic species or isotopes 
\cite{Taglieber2008qdt,Elliott2023qgm}.

\section*{From few- to many-body physics}

The behaviour of quantum mixtures and their ground state configuration depend on many experimental parameters and primarily on the strength of the intercomponent interaction, encoded in the scattering length $a_{12}$ (unless otherwise specified, we always refer to $s$-wave interactions, being the most relevant ones in the ultracold regime). In order to provide a  simple, qualitative picture in which bosonic and fermionic mixtures are compared, let us restrict for simplicity to the case of two gases at zero temperature in a box potential, with equal number of particles ($N_1=N_2$) and equal masses for the two constituents.

The competition between the intercomponent interaction energy ($E_{12} \propto a_{12}$), which can be experimentally tuned through a Feshbach resonance (FR)~\cite{Chin2010fri} (see Supplementary Information), and the energies of the single components ($E_{1},E_2$) can give rise to different configurations of the mixture. In order to obtain a unified picture, we can define the dimensionless parameter $\gamma=E_{12}/\sqrt{E_1 E_2}$ and normalize the total energy of the mixture, $E$, as $\epsilon=E/\sqrt{E_1 E_2}$. For the FF case, since there are no interactions between identical fermions due to the Pauli principle, the intracomponent energies are given by the Fermi energies of the single components, $E_\text{F}$, which are equal in a balanced gas. The Fermi energy introduces a length scale given by the inverse of the Fermi momentum $k_{\text{F}} = (6\pi^2 n)^{1/3} = \sqrt{2m E_{\text{F}}}/\hslash$, with $n$ the density of a single component, so that $\gamma \sim k_\text{F}a_{12}$ and compares the intercomponent scattering length to the mean interparticle distance. For BB mixtures, since intracomponent interactions can be present, we can distinguish between two regimes: 
When all scattering lengths are small and similar in magnitude, $\gamma$ reduces to $ a_{12}/\sqrt{a_1 a_2}$, highlighting the competition between the intra- and intercomponent interactions; otherwise, when the intracomponent scattering lengths are negligible with respect to the intercomponent one, it is convenient to introduce, as for the FF case, an energy scale $E_n$ connected with the inverse of the interparticle spacing $k_{n} = (6\pi^2 n)^{1/3} = \sqrt{2m E_{n}}/\hslash$, so that $\gamma \sim k_na_{12}$.

\begin{figure}[t!]
        \centering
        \includegraphics[width= 0.8
        \linewidth]{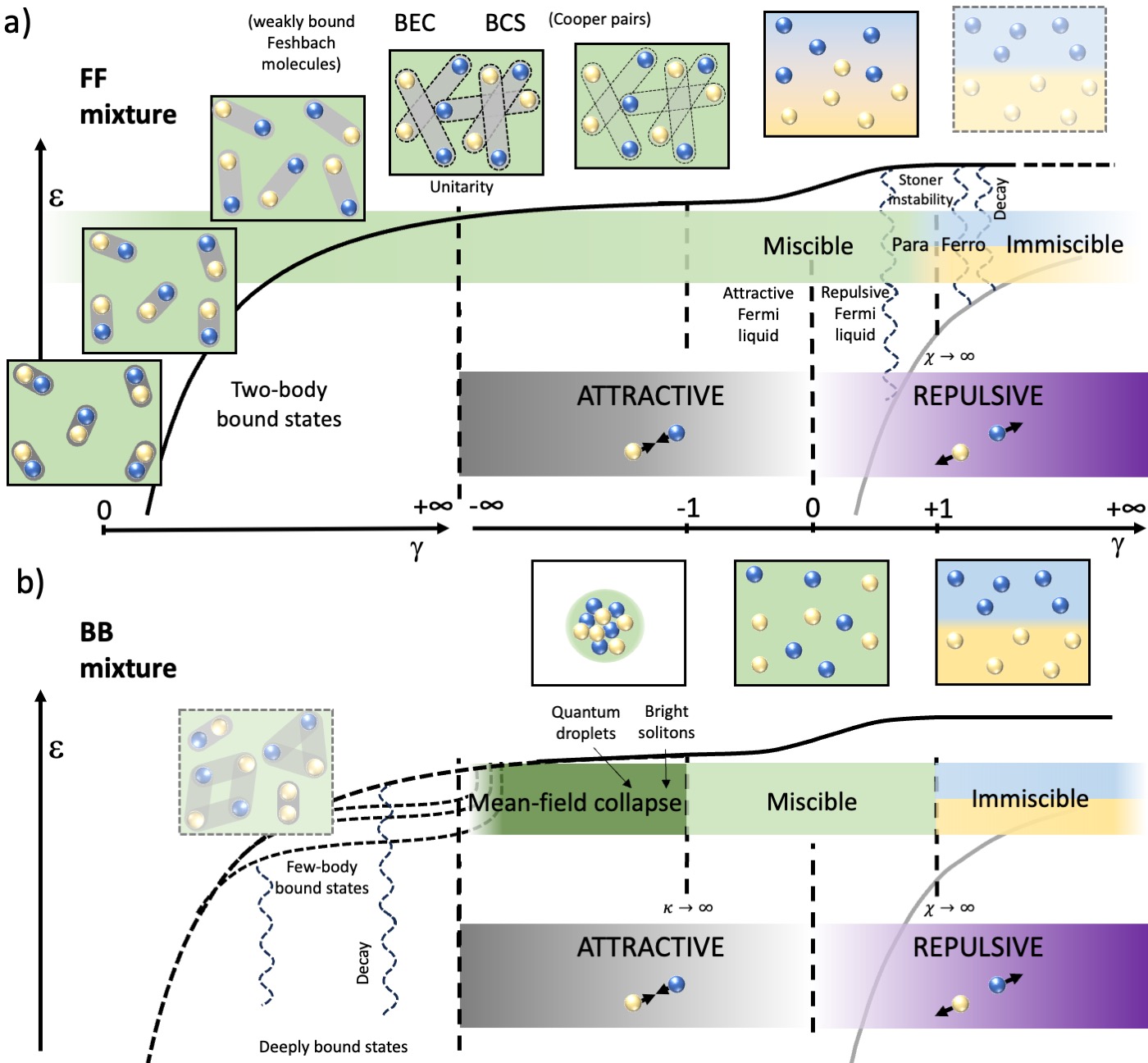}
        \caption{{\textbf{Fermi—Fermi (FF) versus Bose—Bose (BB) mixtures.} Different configurations of FF (a) and BB (b) mixtures for different intercomponent interaction strengths and energies. The positive side of $\gamma$ (dimensionless parameter given by the ratio between the intercomponent energy and the mean intracomponent one)
        is reported twice to follow the energy behaviour across the sign changing of the intercomponent interaction both crossing a Feshbach resonance, with diverging interaction, and crossing the point where the interaction is zero.
        Dashed lines correspond to unstable configurations. 
        Note that in panel (b) only repulsive intracomponent interactions ($a_{1} (a_{2}) > 0$) are considered.}}
        \label{Fig:PhaseDiagram}
    \end{figure}

Figure~\ref{Fig:PhaseDiagram} qualitatively illustrates different possible configurations for FF or BB mixtures as a function of $\gamma$. We choose to graphically show the positive region of $\gamma$ next to the entire $\gamma$ range from $-\infty$ to $+\infty$, following the growth of the system energy. This representation allows to understand what happens both for large scattering lengths (as for instance near the pole of a FR) and for small ones (when $a_{12}$ smoothly changes sign across zero).
For a given value of $\gamma$, more configurations are possible, with corresponding different energies, as the grey line shows on the right side of the diagram.

FF mixtures (Fig.~\ref{Fig:PhaseDiagram}a) always occupy the whole available volume, regardless of $a_{12}$, because of the 
Pauli exclusion principle which forbids the spatial overlap of fermions of the same kind. Therefore, if the particle number is fixed, then the density $n$ is fixed and uniform. 
On the left side of the diagram,
the system ground state is a gas of tightly-bound bosonic molecules, connected with the existence of a two-body bound state, made of two different fermions, which can form also in vacuum, and which features a characteristic size 
growing for increasing $\gamma$ 
(see also next section).
As $\gamma$ becomes very large, 
the size of individual molecules 
becomes comparable with the average interparticle distance. 
The zero-temperature ground state of the system smoothly evolves from a Bose-Einstein condensate of 
molecules to a Bardeen-Cooper-Schrieffer superfluid of Cooper pairs (BCS-BEC crossover) \cite{Zwerger2012tbb}.
On the negative $\gamma$ side, such many-body pairs at any small but finite temperature 
lose correlation as $\gamma$ approaches zero.
The FF mixture then smoothly turns from an attractive to a repulsive Fermi liquid as it enters the $\gamma>0$ region again, but at a higher energy with respect to the paired ground state. Repulsive FF mixtures are interesting since, similarly to electrons in transition metals, they may undergo a para-to-ferro-magnetic phase transition for $\gamma\sim1$ \cite{Stoner1933,Duine2005,Pilati2010,Chang2011,Jo2009,Valtolina2017,Amico2018,Scazza2020} -- once the intercomponent repulsion overcomes the Fermi pressure -- thereby providing a clean framework to explore the textbook Stoner’s model of itinerant ferromagnetism \cite{Stoner1933}. 
However, two important features make the physics of repulsive FF mixtures different from that of ferromagnetic materials in the solid state \cite{Massignan2014pdm}. 
First, unlike electrons in metals, where only the total population is fixed, the populations $N_{1}$ and $N_{2}$ of a FF mixture are generally fixed separately, thus ferromagnetism translates into the formation of spatially-separated domains containing only  
particles belonging to one or the other component. Second, any repulsive FF atomic mixture represents an excited branch that lays above the energy of the (paired) many-body ground state, into which the system can decay through inelastic 
recombination processes, which are resonantly enhanced as $\gamma$ is increased \cite{Petrov2003,Pekker2011}: The ferromagnetic instability in ultracold systems thus inherently competes with the pairing one \cite{Petrov2003,Shenoy2011,Pekker2011,Jo2009,Sanner2012,Scazza2017rfp,Amico2018,Scazza2020}.

BB mixtures (Fig.~\ref{Fig:PhaseDiagram}b) behave very differently. If the intracomponent interactions are attractive, the corresponding component collapses. The interesting case comes when both gases have repulsive intracomponent interactions, allowing for a stable configuration.
Let us start from the right hand side of the plot. For large, positive $\gamma$, the intercomponent repulsion dominates over the intracomponent ones and the two gases undergo phase separation occupying different domains in the available volume \cite{Papp2008tmi}. The critical value for $\gamma$ below which the system becomes miscible is $\gamma=+1$ \cite{Stenger1998sdi,Hall1998doc}. The BB mixture stays miscible and stable 
in the range $-1<\gamma<+1$. When $\gamma$ 
drops below $-1$, attractive interactions dominate and the mixture shrinks occupying a smaller volume than the available one. Within a small range of negative $\gamma$ the mixture can form bright solitons (in 1D) \cite{Cheiney2018bst} or quantum droplets \cite{Cabrera2017qld,Semeghini2018sbq,DErrico2019ooq}, and undergo actual collapse and implosion for stronger intercomponent attraction \cite{Petrov2015qms,DErrico2019ooq,Wilson2021doa}.
If intercomponent interactions are changed fast enough from repulsive to attractive across the resonant region, 
the collapse can be overcome and the mixture enters the bound state region forming Feshbach molecules. 
Since BB mixtures are not subjected to 
the Pauli exclusion principle, not only two-body bound states of different atoms are possible, but
also more exotic few-body clusters, such as Efimov states, may form, that involve more atoms of the same species \cite{Naidon2017epa}. 
Correspondingly, in this regime the system stability is strongly reduced by enhanced loss processes, primarily three-body recombination and inelastic dimer scattering, which drastically limit the lifetime of BB mixtures, even in the non-degenerate regime, at strong interactions $|\gamma|\gg$1. 
For this reason, producing Feshbach molecules is much more challenging in the BB than in the FF case, and BB dimers are typically created via fast magnetic-field sweeps across the FR, employing dilute, thermal atomic samples \cite{Papp2006ooh,Lam2022hps}.

The phenomenology exhibited by FB mixtures, not depicted in Fig.~\ref{Fig:PhaseDiagram}, in most cases qualitatively resembles the BB one, both for what concerns the many-body phenomena that can be accessed, and the collisional stability of the system under strongly-interacting conditions \cite{Modugno2002coa,Ospelkaus2006idd,Ospelkaus2006thi,Zaccanti2006cot,Duda2023}. A detailed classification of the ground state configurations of such mixtures can be found, for instance, in Ref.~\cite{Ufrecht2017ccf}. 

In the following, we enter more in detail and focus on three different regimes: few-body physics, related to problems involving two or three atoms of different kinds in vacuum, the physics of impurities, when single (or a few) atoms of one 
component are immersed in a bath of particles of a different kind, and many-body physics, studying the interplay and the dynamics of two macroscopic,  extended quantum gases.

\subsection*{Few-atom systems}
 
Fundamental insight into quantum mixtures 
comes from the characterization of a few interacting atoms in vacuum. Here, we briefly overview some of the main and well-established aspects of the so-called few-body problem, referring the interested reader to Refs. \cite{Braaten2006uif,Chin2010fri,Petrov2012Tfp,Naidon2017epa,Zaccanti2023mif} for a more detailed and extensive description of this broad, cross-disciplinary research field.
Before discussing the properties of three-atom systems and higher-order ensembles, it is useful to recall the features of the simplest two-body system of just two particles A + B. 
We assume that these are confined in a spherical volume of radius $R_{box}$ (although one could employ any kind of trapping potential), and that they mutually interact via a contact-like, short-range potential, fully-characterized by an $s$-wave scattering length, that can be arbitrarily tuned via a FR (see Supplementary Information).
Figure~\ref{FewBody}(a) shows the low-energy spectrum associated with (the relative motion of) the two particles \cite{Busch1998tca,Pricoupenko2004}, plotted as a function of the normalized interaction parameter $R_{box}/a$, which plays  the same role of $1/\gamma$ within the many-body scenario of Fig.~\ref{Fig:PhaseDiagram}.

Interactions markedly modify the atom-atom spectrum, giving rise to two distinct branches, smoothly connected to the ground-state energy $E_0\sim\!1/R_{box}^2$ (gray dashed line) of the non-interacting system for $a$=0, in strong analogy with the corresponding many-body scenario of Fig.~\ref{Fig:PhaseDiagram}. 
In Figure \ref{FewBody}a, the lowest branch (red line) systematically lays below $E_0$, and it thus corresponds to a net A-B attraction. As the resonance pole is approached from the right side, 
the lowest branch energy monotonically decreases and connects to that of a tightly-bound molecule in vacuum (green short-dashed line) for $R_{box}/a \gtrsim$1. Remarkably, this trend offers a practical way to adiabatically convert ultracold atom pairs into ultracold diatomic molecules, upon sweeping the magnetic field 
from above to below the FR pole (moving from right to left along the horizontal axis in Fig.~\ref{FewBody}a)
\cite{Chin2010fri,Koehler2006poc}. Indeed, `magneto-association' of heteronuclear Feshbach dimers -- followed by a coherent optical transfer to their absolute ro-vibrational ground state \cite{Vitanov2017} --  currently represents the most efficient method to realize quantum gases of polar molecules \cite{Ni2008ahp,Liu2019ooi,Valtolina2020,Duda2023,Bigagli2023}, characterized by a sizable electric dipole moment, and thus by strong, long-ranged interactions. 
The upper branch (blue line) lays instead systematically above the non-interacting energy, corresponding to a net physical repulsion, which progressively increases across the resonance region, asymptotically approaching the energy of the first excited state of the non-interacting system (marked by the light gray dashed line) as $a\rightarrow0^-$. 
Interestingly, at the resonance pole $1/a$=0, the energy of both branches is independent on $a$ and scales as $1/R_{box}^2$ such that the two-body system exhibits a \textit{universal} behaviour, in perfect analogy with the many-body case of unitary Fermi gases, see Fig.~\ref{Fig:PhaseDiagram} and next section. 
As $R_{box}\rightarrow \infty$, the above scenario is maintained but the energy spectrum tends to a continuum of scattering states at positive energies, below which a bound molecular level exists for $a>0$ at energy $\epsilon_0= - \hslash^2/(2 m_r a^2)$, see dashed line in Fig.~\ref{FewBody}(a). 
The energy landscape of Fig.~\ref{FewBody}(a) holds for any homo- and heteronuclear A-B combination except for two identical fermions, for which $s$-wave scattering is forbidden by the Pauli principle. 

The scenario drastically changes when a third particle -- say, a second A atom -- is added to the system. 
To gain an intuitive picture of the three-atom problem, let's assume the A majority particles (of mass $M$) to be much heavier than B (of mass $ m\ll M$), such that we can treat the system within the Born-Oppenheimer approximation \cite{Fonseca1979}. As for the textbook case of the $H_2^+$ molecule, one first solves the Schrödinger equation for the light particle in presence of the heavy ones fixed at given A-A distance $R$, and then employs the $R$-dependent eigenenergies as effective potentials for the heavy particles.

Assuming an intercomponent interaction entirely characterized by a scattering length $a>0$ much larger than the van der Waals range $l_{vdW}$ such that a weakly-bound dimer exists [dashed line in Fig.~\ref{FewBody}(a)], for $R\rightarrow \infty$ the light particle will be localized near either of the two heavy ones. The three-body system thus consists of one AB dimer plus an unpaired A atom and, correspondingly, its energy equals the bare dimer binding energy $\epsilon_0$. 
For finite distance, instead, the configuration where the B particle is localized near the “left” A atom gets mixed with the one where it is localized near the “right” one. As in the double-well problem with tunneling, the symmetric and antisymmetric superpositions lead to two  solutions, $\epsilon_+(R)$ and $\epsilon_-(R)$, respectively symmetric and antisymmetric with respect to the permutation of the heavy particles, $\textbf{R} \leftrightarrow -\textbf{R}$. 
Figure~\ref{FewBody}(b) shows the resulting $\epsilon_{\pm}(R)$, normalized to the dimer energy and plotted as a function of the A-A distance $R$ (in units of $a$). 
Despite originating from a short-range A-B potential, both $\epsilon_{\pm}(R)$ are \textit{long ranged}. The light particle, moving back and forth between the heavy ones, allows  particles to feel their mutual interaction over length-scales greatly exceeding $l_{vdW}$, and on the order of the scattering length, $R\sim |a| \gg l_{vdW}$. 
In particular, $V_-(R) \equiv \epsilon_-(R)-\epsilon_0$>0 corresponds to an effective atom-dimer \textit{repulsion}, which grows for decreasing  distance and reaches the three-atom continuum (see shaded gray area in Fig.~\ref{FewBody}(b)) at $R\!=\!a$. In contrast, $V_+(R) \equiv\epsilon_+(R)-\epsilon_0$ is a purely \textit{attractive} potential which, for $R/a\ll 1$, is found to scale as $V_+(R) \sim -(\hslash^2 c^2)/(2 m R^2)$, with the dimensionless constant $c\sim$ 0.567.
Depending on whether the heavy particles are identical bosons or fermions, the three-body wavefunction must be overall symmetric or anti-symmetric with respect to their permutation. This implies that for bosonic A atoms $V_+(R)$ acts on even atom-dimer angular momentum ($L$) channels -- in particular on the $s$-wave one. For heavy fermionic particles, instead, the induced attraction occurs in odd channels, and primarily on the $L=1$ ($p$-wave) one. Analogously, quantum statistics constrains the $V_-(R)$ repulsion to odd (even) partial waves for majority bosonic (fermionic) particles.
As a result, the effective three-body interactions are long-ranged and multi-channel in nature, although arising from a short-range, $s$-wave direct one. 

   \begin{figure}[t!]
        \centering
        \includegraphics[width= 0.9\linewidth]{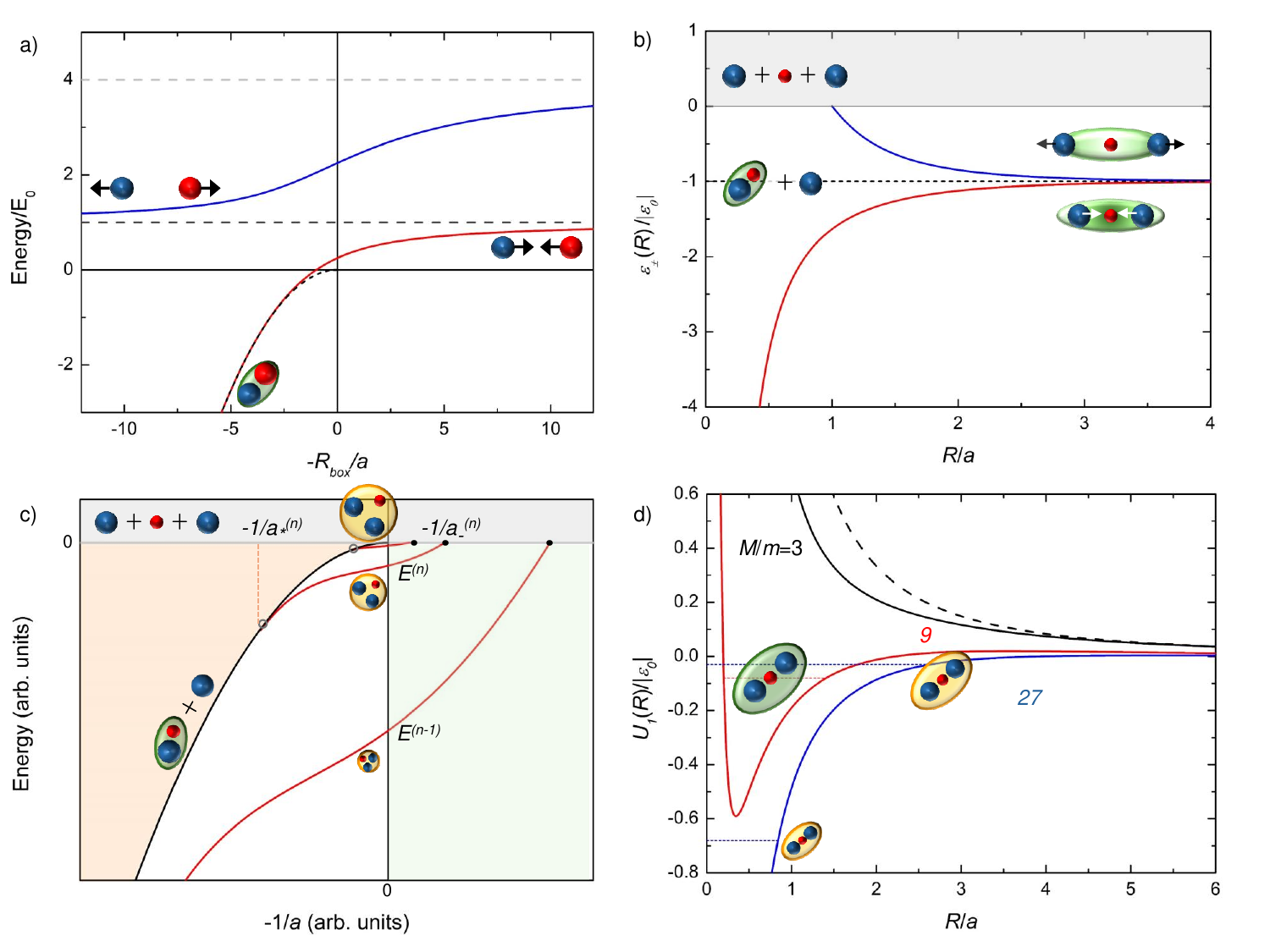}
        \caption{\textbf{The few-atom problem}. (a) Low-energy spectrum of two interacting atoms in a spherical box. (b) Born-Oppenheimer potentials as a function of the distance between the two heavy particles. Red (blue) curve corresponds to an effective induced attraction (repulsion) between the heavy atoms mediated by the exchange of the light one, illustrated by the incoming (outgoing) arrows and the green gradient in the sketch. (c) Sketch of the Efimov scenario. Larger trimer size corresponds to a smaller binding energy. (d) Total Born-Oppenheimer potential $U_1(R)$ for two heavy fermions plus a third atom  as a function of the $M-M$ distance and different $M/m$ values, see legend. Dashed line denotes $V_{cb}(R)$ for $M/m=3$.}
        \label{FewBody}
        \vskip -13pt
    \end{figure}
    
For the case of majority bosonic atoms – and more generally for any system A-B-C of distinguishable particles with at least two resonant pairwise interactions \cite{Naidon2017epa} --  the effect of $V_+(R)$ acting on the lowest partial-wave ($L=0$) channel leads to the celebrated Efimov scenario \cite{Efimov1970ela} (see Refs. \cite{Braaten2006uif,Braaten2007epi,Blume2012,Wang2013,Naidon2017epa} for extensive reviews), schematically illustrated in Fig. \ref{FewBody}(c). 
Connected with the peculiar $1/R^2$ scaling of the induced attraction \cite{Landau1977book} -- that for $1/a=0$ extends from $R \sim l_{vdW}$ up to infinite distances, see Fig. \ref{FewBody}(b) -- the system exhibits an infinite ladder of trimer states ${E^{(n)}}$, characterized by a peculiar “discrete scale invariance”: starting from the most tightly-bound trimer, at energy $E_0\sim \hslash^2/(2 m l_{vdW}^2)$ set by the short-range properties of the A-B interaction \cite{Naidon2017epa}, Efimov states accumulate towards zero-energy according to a geometric series $E^{(n+1)}/E^{(n)}=e^{-2 \pi/s_0}$, where the positive dimensionless constant  $s_0$ depends on the specific system considered,  in particular upon the mass ratio $M/m$  \cite{Naidon2017epa}. 
Moving out of resonance, the energy of each trimer (red lines in Fig.~\ref{FewBody}(c)) is modified in a surprising fashion: for $a<0$, where no dimer exists, the $n$-th Efimov trimer persists, within a kind of 'Borromean binding', up to a critical value $a^{(n)} _-<0$ (see black dots in Fig.~\ref{FewBody}(c)) where it hits the three-atom continuum.
Moving towards small $a>0$ values, while the AB dimer becomes increasingly more bound (black line), the trimer energies progressively approach the dimer one, hitting it at specific $a^{(n)} _*>0$ values (see empty gray circles). Similarly to the ${E^{(n)}}$ energy spectrum at resonance, both series of ${a^{(n)} _-}$ and ${a^{(n)} _*}$ feature the peculiar discrete scale invariance $a^{(n+1)}_-/a^{(n)}_-=e^{\pi/s_0}$. An important consequence of the induced $1/R^2$ attraction characterizing the Efimov's scenario is that it inherently favours the three atoms to approach each other at short distances, $R\sim l_{vdW}$. This favours inelastic loss processes -- such as three-body recombination or inelastic atom-dimer scattering towards deeply-bound molecular levels -- that are enhanced as the resonant regime, or an Efimov resonance, is  approached, limiting the lifetime of $N$>3 boson systems at strong interactions \cite{Rem2013lot,Fletcher2013soa}.
Efimov's prediction \cite{Efimov1970ela}, dating back 1970, has been successfully explored  -- following the first observation of Efimov trimers in a single-component gas of $^{133}$Cs bosonic atoms \cite{Kraemer2006efe} -- in different homo- and hetero-nuclear  systems \cite{Naidon2017epa}. Particularly relevant for the topic of this  review is the exploration of the Efimov scenario in three-component spin mixtures of ultracold $^6$Li atoms \cite{Ottenstein2008cso,Lompe2010ads,Lompe2010rfa,Huckans2009tbr,Williams2009efa}, in $^{41}$K--$^{87}$Rb Bose-Bose \cite{Barontini2009ooh} and  $^{40}$K--$^{87}$Rb Fermi-Bose  \cite{Bloom2013tou} mixtures and, more recently, also in the two extremely mass-imbalanced Fermi-Bose combinations $^6$Li--$^{133}$Cs \cite{Pires2014ooe,Tung2014gso} and $^6$Li--$^{87}$Rb \cite{Maier2015era}. In particular, the latter two systems are significantly more advantageous than homonuclear ones to reveal the presence of few-body clusters, and to test their discrete scale invariance. This can be  understood by considering that for $M/m\gg$1, within the Born-Oppenheimer approximation, $s_0 \propto \sqrt{M/m}$, implying that the larger the $M/m$ ratio, the smaller the scaling factor $e^{\pi/s_0}$, and thus the denser the Efimov spectrum will be \cite{Naidon2017epa}.  Correspondingly, the chance to find, within realistically-accessible ranges of scattering lengths below the unitary limit \cite{Rem2013lot,Fletcher2013soa}, multiple  $a^{(n)}_-$ and $a^{(n)}_*$ resonances is significantly enhanced for highly-imbalanced mixtures, with respect to the equal-mass case. For instance, exact calculations (see Ref. \cite{Naidon2017epa} for details) foresee for $M_{Cs}/m_{Li}\sim$22.1 a scaling constant $e^{\pi/s_0} \sim$4.88, significantly smaller than the one holding for three identical bosons, of about 22.7.  This has allowed for the observation of up to three consecutive Efimov resonances in Li-Cs mixtures \cite{Tung2014gso}, whose relative location  quantitatively confirmed the theory predictions.

When majority fermions, rather than bosons, are considered, the scenario qualitatively changes. On the one hand, the fact that the lowest $s$-wave channel of the three-atom system is associated with the $V_-(R)$ repulsion prevents atoms to approach at short distances,  making resonantly-interacting FF mixtures much more stable than those involving (at least one) bosonic species, e.g. enabling the realization of the unitary Fermi gas and the study of crossover superfluidity \cite{Zwerger2012tbb}. On the other hand, the induced attraction $V_+(R) \propto -1/(m R^2)$ occurs in the $p$-wave channel, where it competes with the effective repulsion set by the centrifugal barrier $ V_{cb}(R) \propto +1/(M R^2)$. This results in an overall potential $ U_{1}(R)= V_+(R)+V_{cb}(R) $ which, as illustrated in Fig.~\ref{FewBody}(d), strongly depends upon the system mass ratio at the qualitative level, contrarily to the previously discussed bosonic case. For $M/m \geq$13.6 (see e.g. blue line for $M/m$=27) $V_+(R)$ dominates over $V_{cb}(R)$ at all distances, leading to the Efimov scenario (see Fig.~\ref{FewBody}(c)), although here trimers carry a non-zero angular momentum. As the mass ratio is progressively lowered below such a critical value, the centrifugal barrier overcomes the induced attraction at short distances, and $U_{1}(R)$ first develops a shallow minimum at large inter-particle distance $R \sim a$ (see red line), that then disappears until $U_{1}(R)$ asymptotically approaches the centrifugal barrier for small mass ratios (see e.g. dashed versus solid black line in Fig.~\ref{FewBody}(d)). Yet, for intermediate mass ratios 8.17$\leq M/m \leq$13.6, the potential well in $U_1(R)$ is deep enough to support (at most two) bound states \cite{Kartavtsev2007let}. Such Kartavtsev-Malykh (KM) trimers starkly differ from Efimov ones: 
Like Feshbach dimers, they are loosely-bound cluster states that exist only for $a>0$ and exhibit universal properties, solely determined by the scattering length and the mass ratio $M/m$, as long as finite-range corrections are negligible \cite{Endo2012}. Moreover, KM trimers (and higher-order non-Efimovian clusters) are expected to be collisionally stable -- since the potential barrier prevents particles to approach each other at short distances, see red line in Fig.~\ref{FewBody}(d). This makes them appealing also at the many-body level, where they are predicted to promote qualitatively new regimes of strongly-correlated fermionic matter \cite{Endo2016,Liu2023} beyond the equal-mass scenario of Fig.~\ref{Fig:PhaseDiagram}(a). Such few-body states lack experimental observation thus far, although the strong $p$-wave atom-dimer attraction revealed in $^6$Li--$^{40}$K mixtures \cite{Jag2014ooa} is a precursor of a KM trimer at sub-critical $M_{K}/m_{Li} \sim$6.6 value. In the future, stable KM trimers may be accessible with newly-realized $^6$Li$^{53}$Cr Fermi mixtures \cite{Ciamei2022,Ciamei2022A,Finelli2024ulc}
($M_{Cr}/m_{Li} \sim$8.8) and other FF systems of smaller mass ratio \cite{Ravensbergen2020rif,Soave2023otf}, but confined in low dimensions (2D, 1D), where the centrifugal barrier is weakened, thereby favouring cluster formation at lower $M/m$ values, see e.g.  \cite{Kartavtsev2009,Pricoupenko2010}.
Finally, we remark that a variety of few-body cluster states with $N$>3 have been also considered and partly explored \cite{Naidon2017epa}. Quite generally, for majority-boson systems one expects that binding of arbitrarily large clusters is allowed by the $1/R^2$ shape of the induced attraction. Although calculations become increasingly difficult with larger number size, this trend has been experimentally-confirmed at least for 4-body Efimovian states \cite{Ferlaino2009efu}. In contrast, binding arbitrarily large numbers of fermions is ruled out beyond five-body clusters, owing to the $p$-wave symmetry of the channel supporting the induced attraction~\cite{Bazak2017}.

\subsection*{Atom-gas (impurity problem)}
By increasing the number of scattering partners that a particle can have, we enter the many-body regime~\cite{Wenz2013fft} and it is convenient to think at such system as an impurity B atom embedded into a continuous, host bath formed by the majority A particles.
Understanding this extremely imbalanced regime is relevant since the impurity limit exhibits some of the critical points of the full zero-temperature phase diagram \cite{Parish2007}.
For weak interactions, the energy of a single impurity is modified with respect to the non-interacting value by the mean-field energy, which is negative for $a_{12}<0$ and positive for $a_{12}>0$. Analogously, the energy of the $n$-mer is shifted by the mean-field effect. The resulting states are the analytic continuation of the vacuum ones described in the previous section (see Fig.~\ref{FewBody}a).
For strong interaction, the impurities will create excitations in the bath, in the form of Bogoliubov modes or particle-hole fluctuations in a Bose condensed or in a degenerate Fermi bath, respectively, which
markedly modify the impurity properties, relative to the bare ones in vacuum.
In this context, it is convenient to describe the system in terms of \textit{quasiparticles} following the concepts originally introduced by Lev Landau and Solomon Pekar \cite{Landau1948emo}, and Herbert Fr\"ohlich \cite{Froehlich1954eil}. In this framework, an electron moving through a ionic crystal can be conveniently described as a quasiparticle, formed by the bare electron dressed by the polarization cloud induced by its interaction with the lattice ions (see Fig.~\ref{fig:polaron}a). 
Similarly, the complicated many-body system, realized by letting few impurity atoms strongly interact with a (Fermi or Bose) gas, can be mapped into a weakly-interacting liquid of quasiparticles, made by the bare impurities dressed by excitations of the host bath, with renormalized properties, such as energy and effective mass. In this picture, the overlap between the quasiparticle and the bare particle is encoded in the so-called \textit{quasiparticle residue} \cite{Massignan2014pdm}.
The literature refers to these quasiparticles as \textit{polarons, dimerons, trimerons}, etc.,  when arising from the dressing of a single B impurity, a A-B molecule, a A-A-B trimer, and so on  \cite{Lan2014asi}.
In particular, the dressed state connected to the atomic impurity attractively (repulsively) interacting with the host gas is referred to as the attractive (repulsive) polaron (see Fig.~\ref{fig:polaron}b and c). 
Note that the dimeron, also referred to as \textit{dressed molecule} or \textit{molaron}, should not be confused with the bare \textit{Feshbach molecule} (see green line in Fig.~\ref{FewBody}a): like for the polaron case, its energy is strongly modified by the surrounding medium and, as discussed below, it may, or may not, represent the system ground state.

\subsubsection*{Fermi and Bose polarons}

\begin{figure}[t!]
        \centering
        \includegraphics[width= 1\textwidth]{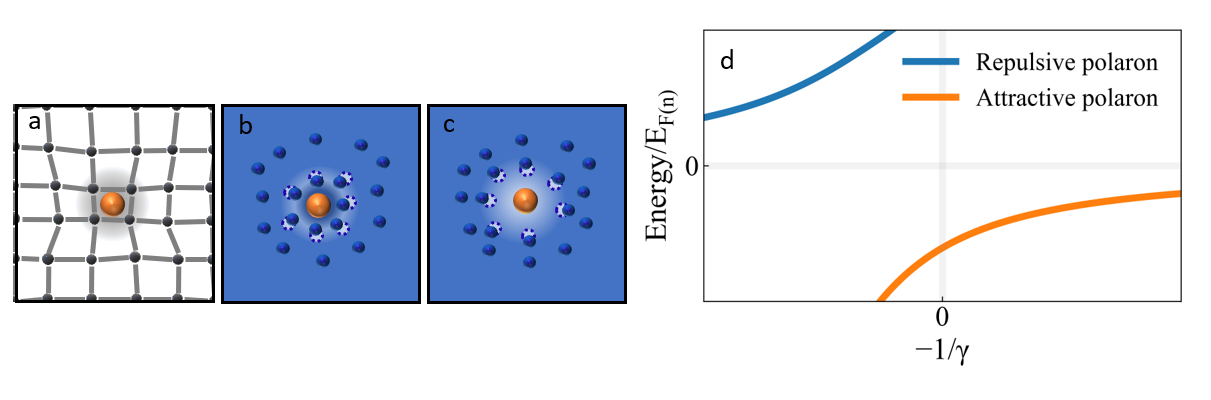}
        \caption{\textbf{Polarons.} Pictorial representation of a polaron in solid state (a) and its counterpart in quantum gases in the attractive (b) and repulsive (c) cases. d) Energy spectrum showing the repulsive and attractive polaron branches
        (numerical data for a 
        homonuclear spin mixture, courtesy of P. Massignan).}
        \label{fig:polaron}
        \end{figure}

While the statistics of the single impurity is irrelevant,
the one of the bath defines the polaron properties, and we distinguish between Bose and Fermi polarons, depending whether the bath is formed by a degenerate Bose or Fermi gas, respectively.
The Bose polaron is analogous to polarons created in solid state systems described by the Fr\"ohlich Hamiltonian \cite{Grusdt2016nta}, whereas the Fermi polaron is a prototypical
realization of the fundamental building block of a Landau’s  Fermi liquid \cite{Massignan2014pdm}.

In Fig.~\ref{fig:polaron}d the energy landscape of the impurity-bath system is sketched as a function of the intercomponent interaction.
Contrarily to the two-body system of Fig.~\ref{FewBody}a, here three rather than two distinct spectral features are identified: one corresponding to the attractive polaron (orange line), one to the metastable repulsive polaron (blue line) \cite{Scazza2022rfa} and a third dimeronic one, not shown, asymptotically connected, for vanishing bath density, to the bare Feshbach molecule presented in Fig.~\ref{FewBody}a.
While this scenario qualitatively holds for both bosonic and fermionic media, some caveats are necessary.
First of all, in the case of a bosonic bath, the dimeron energy asymptotically approaches the attractive polaron branch from above while never crossing it, such that the attractive polaron is the system ground state at all interaction strengths. 

In the case of a fermionic bath, instead, the two branches intersect at the so-called \textit{polaron-to-molecule transition} where the ground state changes from the attractive polaron to the dimeron.

Experimentally, it has been observed that this crossing point is smoothed by increasing impurity concentration and temperature \cite{Ness2020ooa}.  
Secondly, the stability of the bath plays a crucial role in determining the quasiparticles lifetime. 
Thanks to the Pauli exclusion principle, the Fermi bath is stable against $n$-body losses at all interaction strengths and, therefore, attractive polarons (and dimerons) are found to persist over timescales exceeding  by several orders of magnitude the typical Fermi time of the host gas \cite{Schirotzek2009oof}. In contrast, enhanced inelastic processes that affect bosonic media in the resonant regime (see previous section), strongly reduce the quasiparticles lifetime, relative to their fermionic counterparts. 
In the Fermi case \cite{Schirotzek2009oof, Nascimbene2009coo} the impurity properties under resonantly-interacting conditions appear to share universal behaviour, solely controlled by the bath density -- hence by $E_F$ -- in strong analogy with both the two-particle system (see Fig.~\ref{FewBody}a and related discussion) and the unitary Fermi gas case \cite{Zwerger2012tbb}, realized with balanced FF mixtures, see next section. In the Bose case, instead, due to three-body correlations, the Bose polaron is generally not scale invariant at unitarity, but depends on an additional Efimov length scale \cite{Levinsen2015iia}. Nonetheless, some universal behaviours have been observed in experimental works on Bose polarons \cite{Yan2020bpn, Etrych2024}. 

Introduction of a mass asymmetry between the impurity atom and the particles of the host gas can qualitatively change the impurity problem \cite{Massignan2014pdm,Christianen2024pdf}. In particular, light impurities are expected to promote the emergence of novel types of quasiparticles within fermionic media \cite{Mathy2011tma,Liu2022}, linked to the existence of higher-order clusters at the few-body level discussed in the previous section. Similarly, light impurities in a Bose gas are sizably affected by increased three-body correlations even at weak coupling and, for stronger intercomponent attraction, the atom-like polaronic state is predicted to smoothly cross-over to a (dressed) Efimov trimer \cite{Levinsen2015iia} or to decay into large Efimov clusters \cite{Christianen2022}. In the opposite limit of an infinitely massive impurity, in a fermionic bath the interacting many-body system is predicted to completely lose any overlap with the non interacting one \cite{Goold2011oca,Knap2012tdi}, leading to the so-called Anderson orthogonality catastrophe \cite{Anderson1967ici}. An analogue effect is predicted also for impurities in a non-interacting Bose gas, for both infinite and finite impurity mass \cite{Guenther2021mii}.

Standard experimental techniques to investigate polaronic systems are provided by spectroscopic methods \cite{Vale2021spo}. 
The energy landscape is routinely accessible via radio-frequency spectroscopy, which led to the first observation of the attractive Fermi polaron in an ultracold $^6$Li FF spin mixture \cite{Schirotzek2009oof}. The full polaron spectrum was recorded for the first time in a FF heteronuclear mixture of $^6$Li and $^{40}$K \cite{Kohstall2012mac}, observed then also in a $^6$Li FF \cite{Scazza2017rfp} and in a $^6$Li--$^{41}$K FB mixture\cite{Fritsche2021sab}. The Bose polaron spectrum was recorded for the first time in parallel in a heteronuclear mixture of $^{87}$Rb--$^{40}$K \cite{Hu2016bpi} and a spin mixture of $^{39}$K \cite{Jorgensen2016ooa}.
The out of equilibrium formation dynamics of the polaron has been investigated via Ramsey interferometric techniques for both the Fermi \cite{Cetina2016umb} and the Bose \cite{Skou2021ned} polaron.
While other polaronic quantities can be inferred from radio-frequency spectroscopy, such as the effective mass \cite{Scazza2017rfp} and the polaron residue \cite{Schirotzek2009oof}, there are more suitable tools, such as Rabi oscillations \cite{Kohstall2012mac, Scazza2017rfp, Adlong2020qlo}, transport measurements \cite{Chikkatur2000sae,Nascimbene2009coo}, Raman \cite{Ness2020ooa} and momentum-resolved photoemission \cite{Koschorreck2012aar} spectroscopy. 

By increasing the number of impurities, interactions between the corresponding polarons, mediated by modulations of the bath, appear. While the quantum statistics of the impurity is irrelevant in the single impurity limit, it is crucial for a finite impurity concentration: in particular the mediated interactions between polarons are always attractive or always repulsive for bosonic or fermionic impurities, respectively, regardless of the sign of the interaction between the impurities and the bath \cite{Yu2012iii}. Such interactions were experimentally observed for polarons formed by both bosonic and fermionic impurities in a Fermi sea in Ref.~\cite{Baroni2023mib}. 
Additionally, due to the attractive nature of the effective interaction between two polarons formed by bosonic impurities, the formation of bound states of two polarons, named \textit{bipolarons}, has been theoretically predicted \cite{Camacho2018bia} but not yet experimentally observed. 

We conclude this section remarking that nowadays the investigation of the impurity problem \cite{Mistakidis2022poi} is, on the one hand, not only restricted to 3D systems, but extends also to reduced dimensionality, such as 2D \cite{Koschorreck2012aar} and 1D \cite{Palzer2009qtt,Catani2012qdo,Meinert2017boi}, on the other hand, not only restricted to neutral mixtures with contact interactions, but also encompasses charged impurities \cite{Perez2021cha, Zipkes2010ats} 
or Rydberg atoms \cite{Camargo2018cor}.

\subsection*{Gas-gas (many-body physics)}

Since FF and BB mixtures are usually investigated in different interaction regimes and the physics of the many-body system strongly depends on the quantum statistics of the constituents, here we treat the two types of mixtures separately. For a general overview of many-body physics investigated with ultracold atoms, considering also lattices and lower dimensionality, we refer the reader to Ref.~\cite{Bloch2008mbp,Gross2017qsw}.

\subsubsection*{Fermi-Fermi}

Let us now consider a mixture of fermions with equal population and equal masses, such as the homonuclear spin mixture sketched in Fig.~\ref{Fig:PhaseDiagram}. 
Because of Pauli blocking there are no intracomponent interactions and the main interest in such mixtures arose from the possibility to investigate a strongly interacting gas without being affected by strong losses, contrary to the BB case. In particular, thanks to Feshbach resonances, the intercomponent interactions can be tuned from repulsive to attractive, crossing the pole of the resonance, where scattering length $a_{12}$ diverges and the system exhibits \textit{universal} features~\cite{Hu2007uto}.

\begin{figure}[t!]
        \centering
        \includegraphics[width= 0.5\textwidth]{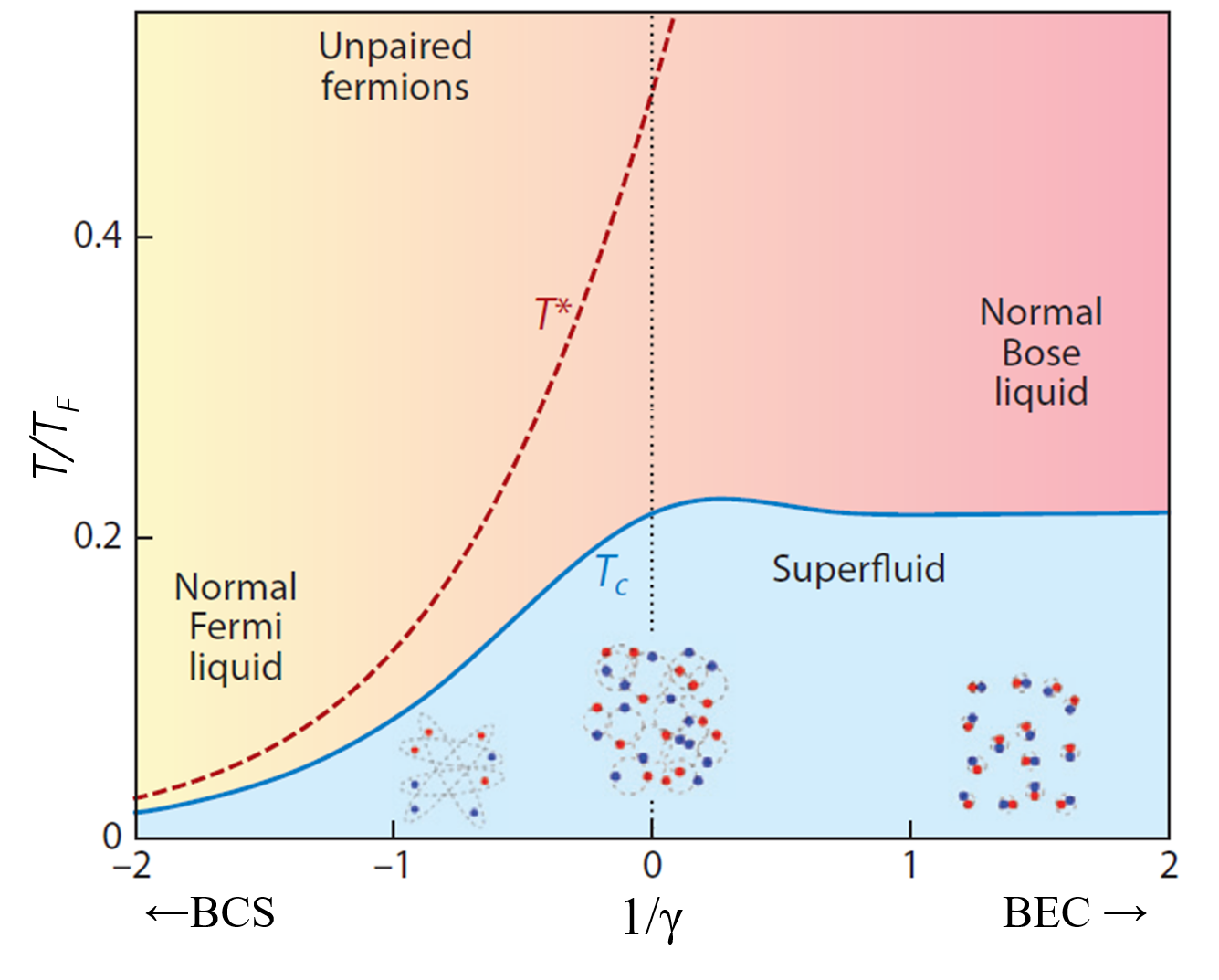}
        \caption{Phase diagram of the balanced mixture. The red-dotted and blue continuous lines indicate the temperatures below which the pairs are formed, $T^*$, and condensed, $T_c$, respectively. Adapted from \cite{Randeria2014cfb}. 
        }
        \label{fig:pairing}
        \end{figure}

The behaviour of a balanced attractive Fermi-Fermi mixture across a Feshbach resonance is captured by the so-called BEC-BCS crossover~\cite{Zwerger2012tbb}, the phase diagram of which is reported in Fig.~\ref{fig:pairing}. 
As already pointed out in the discussion of the few-body physics, for positive scattering lengths $a_{12}$ there exists a bound state, which is more bounded the smaller the scattering length. 
If now we consider a macroscopic number of atoms we have a large number of such molecules which, being created by bounding two fermions, are bosonic in nature and so can undergo Bose-Einstein condensation for temperatures below the critical temperature $T_c$, of the order of fractions of the Fermi temperature $T_\text{F}$: the mixture behaves as a superfluid of a single bosonic component. We refer to this interaction regime as the BEC side of the resonance. Condensed molecules in this regime have been first observed in homonuclear mixtures of $^6$Li~\cite{Jochim2003bec, Zwierlein2003oob} and $^{40}$K~\cite{Greiner2003eoa, Regal2003cum}.

For negative scattering lengths, instead, two atoms can form a large pair with total momentum equal to zero, in analogy with the Cooper pairs of electrons predicted by the BCS theory of superconductivity \cite{Bardeen1957tos}, and we identify this side of the resonance as the BCS side.
The fact that the Cooper pairs have zero momentum arises from the fact that for equal populations and equal masses the Fermi surfaces relative to the two components are equal. One can see that pairing of atoms with opposite momentum is favoured because of their higher density of states leading to a higher binding energy with respect to pairs formed with not zero momentum \cite{Cooper1956bep}.
If there is instead a population or a mass imbalance in the mixture, the two Fermi surfaces do not match anymore and the system, as we shall see, has to find a way to compensate this mismatch. 
Contrary to the molecular states on the BEC side, the Cooper pairs do not exist in vacuum and are the result of the many-body nature of the gas. Indeed, in a Fermi sea, because of Pauli blocking, we can describe the system considering only atoms whose momentum lies on top of the Fermi surface, making the gas an effectively 2D system, for which a loosely bound state exists for infinitesimally small attraction.  These pairs can also condense for low enough temperature, but now the critical temperature scales exponentially with the inverse of the scattering length, as can been seen in the phase diagram in Fig.~\ref{fig:pairing}. Observation of Cooper pairing in momentum space has been recently observed in a mesoscopic two-dimensional Fermi-Fermi mixture~\cite{Holten2022ooc}.
For scattering lengths close to the pole of the Feshbach resonance, there is a crossover between a superfluid of condensed molecules and one of condensed Cooper pairs, where the pair is not as bounded as a molecule and not as loose as a Cooper pair. 
Condensation of such pairs was first observed in Ref.~\cite{Regal2004oor} and the superfluid nature of the gas across the resonance was investigated in \cite{kinast2004efs,Bartenstein2004ceo, Bourdel2004eso, Zwierlein2005vas}. A review of early experimental achievements in the BEC-BCS crossover can be found in Ref.~\cite{Grimm2007ufg_chap}.

Let's consider now an unbalanced mixture and let's start with the case of population imbalance and equal masses, for simplicity.
As we saw until now, in the case of a balance mixture we can have a superfluid formed by pairing all the atoms composing the gas, in the limit of one single impurity in the Fermi sea of a majority of atoms, a polaron is formed. There should be then a phase transition between these two states, connecting the superfluid to the normal Fermi gas \cite{Lobo2006nso}. In particular,  superfluidity breaks down at the so-called Chandrasekhar–Clogston limit \cite{Chandrasekhar1962ano, Clogston1962upf}, where the difference between the two Fermi energies, arising from the population imbalance, is larger than the energy gained in creating a pair. Experimentally, the critical population imbalance for which superfluidity is lost has been investigated in Refs.~\cite{Zwierlein2006fsw, Zwierlein2006doo, Partridge2006pap}.

Superfluidity in an imbalanced Fermi gas can appear in different ways, depending on how the system compensates the mismatch in the Fermi energies of the two components. For detailed reviews on imbalanced Fermi gases, we refer the reader to Refs.~\cite{Chevy2010ucp, Gubbels2013ifg}

In the simplest scenario condensed pairs are created, which phase-separate with the remaining unpaired atoms, resulting in two spatially separated phases given by a superfluid and a normal \textit{polarized} Fermi gas created by the unpaired atoms, as observed in Ref.~\cite{Zwierlein2006fsw}.

More exotic superfluid phases are predicted \cite{Chevy2010ucp}, even though not yet experimentally observed. Two main scenarios are possible: if pairing occurs between particles at the surface of their respective Fermi seas, the pairs acquire a momentum equal to the difference of the two Fermi momenta, and which can be interpreted as a spatial modulation of the order parameter. This is named Fulde–Ferrell–Larkin–Ovchinnikov (FFLO) phase \cite{Combescot2007itf, Kinnunen2018tff, Pini2023eoa}. Another possibility is to form pairs at zero momentum at the cost of opening a gap inside the Fermi sea of the majority component, Sarma phase~\cite{Sarma1963oti, Gubbels2006spi}. 

It is theoretically predicted that adding mass imbalance to a system would rise the critical temperature for these exotic pairings, favoring the elusive experimental observation of such states \cite{Baarsma2010pam, Pini2021bmf}. Nowadays available mass imbalanced FF mixtures are $^6$Li--$^{40}$K~\cite{Grimm2023fqm, Voigt2009uhf}, $^{161}$Dy--$^{40}$K~\cite{Grimm2023fqm}, $^6$Li--$^{173}$Yb \cite{Hara2011qdm, Green2020fri}, $^6$Li--$^{53}$Cr \cite{Ciamei2022,Finelli2024ulc}, and $^6$Li--$^{167}$Er \cite{Schaefer2023oof}.

As anticipated in the previous sections, also repulsive FF mixtures have attracted a growing interest that, for bulk systems that are the subject of this work, mainly connects to the possible exploration of the textbook Stoner's model for itinerant ferromagnetism \cite{Stoner1933,Duine2005,Pilati2010,Chang2011}, see Ref. \cite{Massignan2014pdm} for an extensive review. This arises from the fact that such ultracold systems embody the two ingredients of Stoner’s Hamiltonian \cite{Stoner1933} – Fermi pressure and short-range repulsion – free from intricate band structures, additional kinds of interactions, and disorder inherent to any condensed matter system.
However, experimental attempts in this direction revealed a much richer and more subtle behaviour, very different from the original Stoner scenario. 

Although convincing signatures for a ferromagnetic instability within the repulsive Fermi liquid
have been obtained, through studies of spin dynamics of a  FF mixture prepared in an artificial magnetic domain-wall structure \cite{Valtolina2017}, and via
time-resolved quasi-particle spectroscopy both on balanced \cite{Amico2018} and imbalanced \cite{Scazza2017rfp} mixtures, already early experiments \cite{Jo2009,Sanner2012} found the system dynamics to be fundamentally affected by another type of instability, antithetical to ferromagnetism, and associated with the tendency of repulsive fermions to combine into weakly bound pairs. This latter mechanism is intimately linked to the short-ranged nature of the interatomic interaction: the strong repulsion, $\gamma \sim$1 (see Fig.~\ref{Fig:PhaseDiagram}(a)), necessary for ferromagnetism to develop \cite{Stoner1933,Duine2005,Pilati2010,Chang2011} can only be attained if a weakly bound molecular state exists below the two-atom scattering threshold, see Fig.~\ref{FewBody}(a). As such, the repulsive Fermi gas represents an excited metastable branch of the many-body system, inherently affected by decay processes towards lower-lying molecular states, which become faster for larger repulsive interactions \cite{Petrov2003,Shenoy2011,Pekker2011,Sanner2012}.
Pump-probe spectroscopic studies \cite{Amico2018} provided  measurements of the rates at which  pairing and short-range ferromagnetic correlations develop after creating a strongly repulsive Fermi liquid, finding the two instabilities to rapidly grow over comparable timescales. Interestingly, the same survey revealed that both mechanisms persist at long times, leading to a semi-stationary
regime consistent with a spatially heterogeneous phase \cite{Amico2018,Scazza2020}: a quantum emulsion wherein paired and unpaired fermions macroscopically coexist while featuring phase segregation at the micro-scale of a few interparticle spacings. It is interesting to remark how a similar spontaneous emergence of spatially inhomogeneous states occurs in strongly correlated electron systems: the simultaneous presence of multiple interaction mechanisms and the concurrence of distinct competing instabilities foster the emergence of nanometer-scale heterogeneous structures that host different phases and order parameters \cite{Dagotto2005,Dagotto2003}. 

While the study of repulsive Fermi mixtures has focused mainly on homonuclear systems, we remark that well-defined repulsive polarons - the building blocks of a repulsive Fermi liquid - were first revealed in mass-imbalanced $^6$Li$^{40}$K FF mixtures \cite{Kohstall2012mac}. It is also important to notice how a strong mass-asymmetry may greatly facilitate the development of macroscopic magnetic domains within a repulsive FF mixture: On the one hand, the reduced Fermi pressure of the heavy component can sizeably reduce the critical repulsion for ferromagnetism to emerge  \cite{Cui2013psi}. On the other hand, the pairing rate strongly depends upon the mass ratio of the two components, and it can be starkly decreased, with respect to the equal-mass case \cite{Petrov2003}.

\subsubsection*{Bose-Bose}
\begin{figure}[b!]
        \centering
        \includegraphics[width= \textwidth]{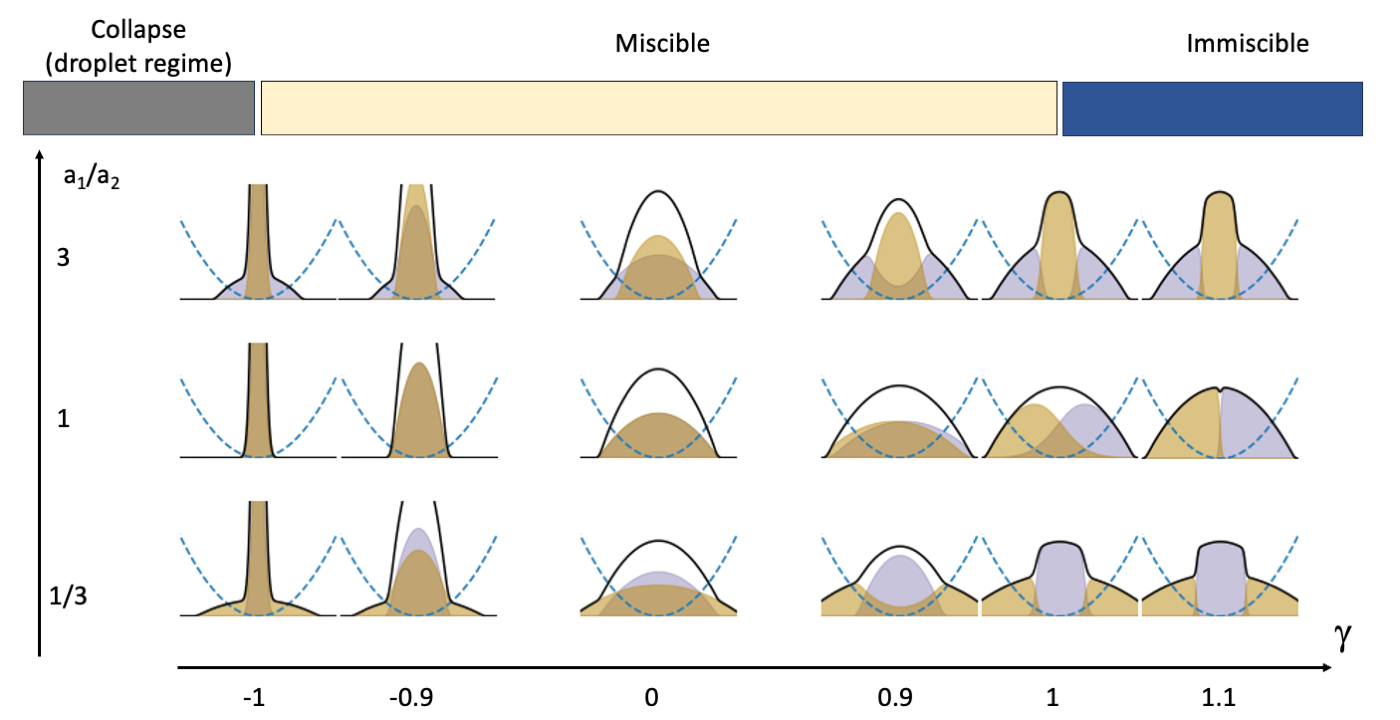}
        \caption{\textbf{Miscibility and buoyancy.} GPE simulation of the density distribution for a BB mixture, with equal masses of the constituents, in a harmonic trap (dashed) for different $a_1/a_2$ and $\gamma$ ($a_2$ is kept fixed). Shaded colored regions illustrate the individual densities, while the solid line shows the total density. A small gradient was added to break the left-right symmetry.}
        \label{fig:miscibility}
\end{figure}

As anticipated while describing Fig.~\ref{Fig:PhaseDiagram}, two condensates can either mix together occupying the whole available volume,  separate in different domains, or collapse \cite{Ho1996bmo}. Since the study of the unitary Bose gas is limited to far-from-equilibrium dynamics \cite{Makotyn2014udo, Eigen2018upd} due to the strong atom losses, we consider here a BB mixture of two independently stable gases (finite and positive $a_1$ and $a_2$) characterized by an intercomponent scattering length $a_{12}$ (positive or negative) of the order of the intracomponent ones, and discuss the possible scenarios in the miscible and immiscible regimes, illustrated in Fig.~\ref{fig:miscibility} for the case of equal masses for the two components. 

In the miscible regime ($a_{12}^2<|a_1a_2|$), if the trapping potential is not flat, the two gases have equal ground state spatial profiles when $a_1=a_2$ \cite{Bienaime2016sdo}, whereas for $a_1 \neq a_2$, buoyancy effects come into play with the least repulsive gas concentrating towards the trap minimum and the other gas redistributing in the outer regions \cite{Hall1998doc} (see examples in Fig.~\ref{fig:miscibility}). 
The many-body mixture configuration and dynamics can be conveniently described introducing the total density ($n=n_1 + n_2$) and their density difference, or magnetization ($m=n_1 - n_2$). These two channels correspond to the in-phase and out-of-phase excitations of the mixture.
Figure~\ref{fig:miscibility} shows examples of density distributions of both components for different $a_1/a_2$ combinations and different $\gamma$, specifically around the critical values $\gamma=\pm 1$,. Note that at $\gamma=0$ each component has a pure Thomas-Fermi profile, with radii associated with their own scattering length. Approaching $\gamma=-1$, the total density is strongly enhanced, consistently with a divergence in the compressibility $\kappa$. As $\gamma$ tends to +1, instead, the total density is barely affected, while the magnetization shows a peak, in agreement with a divergence of the magnetic susceptibility $\chi$.
The specific features of total density and magnetization channels, such as the interaction energy scales, the speeds of sound, and the healing lengths \cite{Lamporesi2023tcs}, depend on the interaction constants and on the densities. In the particular case of equal masses, densities and intracomponent interactions ($a_1=a_2=\bar{a})$, the ratio between the quantities associated with the total density and magnetization channels is a function of $(\bar{a} + a_{12})/(\bar{a} - a_{12})$.
The larger the difference between the two channels, the more they decouple and independent investigations can be implemented in a clean environment \cite{Bienaime2016sdo,Kim2020oot,Cominotti2022oom}. In general, though, complex many-body dynamics emerges, causing damping and oscillation frequency \cite{Wilson2021qdm, Cavicchioli2022ddo}
changes or turbulence \cite{Kim2017css}.

In the immiscible regime ($a_{12}>\sqrt{a_1a_2}$), bosonic mixtures spatially self-organize in different single-component domains in order to minimize the energy cost associated with the strong intercomponent repulsion \cite{Pyzh2020pso}. Beyond-mean-field effects can also lead to equilibrium configurations with bubbles of mixed phases coexisting with a pure phase of one of the two components in the case of either mass imbalance or different intracomponent interactions \cite{Naidon2021mbi}.
Note that immiscibility and magnetization excitations can only be observed in extended systems with a characteristic size that exceeds the magnetization healing length. Such features are instead inhibited in systems smaller than the healing length \cite{Yi2002sma,Frapolli2017sbe}. 

Thanks to buoyancy and immiscibility, quantum mixtures support an extremely rich variety of topological defects such as different kinds of solitons \cite{Kevrekidis2004fom} and vortices, compared to the case of single component gases.
Exotic solitons \cite{BakkaliHassani2021roa,RomeroRos2024ero,Zhang2015ssi}, soliton trains \cite{Hamner2011god}, magnetic solitons \cite{Farolfi2020oom,Chai2020msi}, half quantum vortices \cite{Seo2015hqv} and stable massive multicharged vortices \cite{Richaud2023mdv} are some of the topological structures that cannot be found in single component condensates.

In the special case of homonuclear spin mixtures, the presence of a coherent coupling radiation between two internal states, makes the individual phases lock one to each other and allows for state interconversion (spin is no longer conserved). 
The mixture state can be described with the Bloch vector formalism and treated in analogy with spin dipoles in the presence of an effective magnetic field following Landau-Lifshitz formalism \cite{Farolfi2021qti} in the continuum.
Such a coupled system is also suitable for investigating confinement physics in analogy to high-energy physics, as pairs of half quantum vortices behave in a very similar way to quark pairs \cite{Son2002dwo,Eto2018coh}.
If the difference between the mean intracomponent interactions and the intercomponent one is negative, 
the mixture shows a para- to ferromagnetic- transition that can be studied and used for investigating magnetism in superfluid systems \cite{Zibold2010cba,Cominotti2023fia}
as well as simulating quantum field theory predictions such as false vacuum decay \cite{Zenesini2024fvd} or cosmological curved space-time geometries \cite{TolosaSimeon2022cae}. 
When the states are coupled via a two-photon Raman transition with a non-negligible relative momentum, external and internal degrees of freedom couple (spin-orbit coupling) and the gas behaves as in the presence of synthetic magnetic fields \cite{Lin2009smf,Lin2011soc} and can exhibit supersolid features \cite{Putra2020sco}.

When intracomponent interactions are attractive and stronger than the mean of the intracomponent ones ($a_{12}<-\sqrt{a_1a_2}$), mixtures can either form stable configurations or undergo collapse.
Quantum droplets are self-bound many-body states that exist for small numbers of atoms up to a critical interaction strength, when $a_{12}^2 \sim\! a_1 a_2$. Originally predicted by Dmitry Petrov \cite{Petrov2015qms} as a result of beyond mean-field effects, they were then observed in mixtures of different spin states of $^{39}$K atoms \cite{Cabrera2017qld,Semeghini2018sbq} and in heteronuclear mixtures of KRb \cite{DErrico2019ooq}. 
In one-dimensional geometry also bright soliton solutions, where dispersion compensates for attractive interactions, have been observed \cite{Cheiney2018bst}.

Including the option to have a finite mass imbalance between the two components, new possible phases arise involving bubbles of mixed phases coexisting with a pure phase of one of the two components \cite{Naidon2021mbi}.

\subsubsection*{Fermi-Bose}
As anticipated in the introduction, Fermi-Bose mixtures share many features with the BB ones. 
The first FB mixtures that have been brought to double quantum degeneracy are $^6$Li--$^7$Li~\cite{Truscott2001oof, Schreck2001qbe}, $^{40}$K--$^{87}$Rb~\cite{Roati2002fbq, Goldwin2004mot, Koehl2005fai }, and $^6$Li--$^{23}$Na~\cite{Hadzibabic2003fii}. In Ref.~\cite{Onofrio2016poo} the reader can find a summary of most of the FB mixtures used nowadays and their peculiarities.
We can identify two type of FB mixtures: in one case, the fermionic part can be given by a FF mixture itself, which can be tuned to the superfluid regime, having de facto a BB mixture of two superfluids~\cite{Ferrierbarbut2014amo}. In the other case, the mixture is a genuine two-component Fermi-Bose system, where the fermionic part is given by a polarized degenerate Fermi gas. In this case, similarly to the BB case, the system can undergo phase separation~\cite{Viverit2000zpd, Modugno2007fbm, Lous2018pti} or collapse \cite{Modugno2002coa}, but also novel possibilities can arise, that is, tailoring interactions between the fermionic component thanks to the mediation of the bosonic part \cite{Efremov2002pwc, Kinnunen2018ipw}, or, vice versa, creating long range interactions between bosons mediated by the fermions \cite{DeSalvo2019oof, Edri2020oos, Arguello2022tlr}.

\section*{Conclusions and Perspectives}

In this review article, we provided an introductory overview of (some of) the physical phenomena and main experimental progresses in the research field on ultracold quantum mixtures, from the three relevant perspectives of few-body physics, impurity problems, and many-body physics. 
The wealth of possible states arising in these three different sectors and the intertwining of few-to-many-body physics were discussed by taking into account both the role of a mass asymmetry and distinct quantum statistics of the mixture constituents, which lead to striking differences in the ultracold regime. 

As anticipated in the Introduction, we intentionally restricted our discussion to bulk systems confined in three-dimensional potentials, only marginally mentioning the extremely rich fields of research on ultracold atomic mixtures in optical lattices\cite{Gross2017qsw, Schäfer2020tfq}, reduced \cite{Cazalilla2011odb, Mistakidis2023fbb} or extended (synthetic) \cite{Ozawa2019tqm, Fabre2024atq} dimensionality. 
Additionally, here we focused on the simplest, and most standard case of contact $s$-wave interactions between the two components: We emphasize that a wealth of phenomena and applications, not discussed in this work, are expected to emerge when higher-order (e.g. $p$- or $d$-wave) partial waves~\cite{Cui2018}, or long range interactions -- such as dipolar ones~\cite{Chomaz2023dpa} -- are considered. In this context, it is also relevant to mention the important experimental progress in producing ultracold atom-ion mixtures~\cite{Tomza2019RMP,Lous2022}, and in manipulating their intercomponent interactions\cite{Weckesser2021oof}.

Homonuclear spin mixtures, both bosonic and fermionic, are powerful and highly tunable systems. They made it possible to realize the first experimental superfluid mixture and offer a valuable platform for exploring the complex dynamics of many-body systems.
In this context, interesting directions are offered by further exploration of spinor physics with nontrivial ground state configurations and excitations \cite{StamperKurn2013sbg}. 
Furthermore, coherent coupling between internal states with and without spin-orbit coupling \cite{Recati2022ccm} can be used to generate highly-tunable quantum simulators with which to tackle a variety of phenomena belonging to other fields of physics, such as cosmology and gravitation \cite{Fischer2004qso, Visser2005mkg,Liberati2006aqg,Butera2017bhl}, magnetism \cite{Cominotti2022oom}, and high energy physics \cite{Eto2018coh}. Ultracold spin mixtures can also reveal their high potential for the implementation of spintronic devices with superfluid features, where the spin introduces an extra degree of freedom with respect to atomtronic applications \cite{Amico2022cac}.

Heteronuclear mixtures offer the possibility to engineer species-selective optical potentials for an almost independent manipulation of the different components \cite{LeBlanc2007sso}. Thus we can
control the single species separately in terms of motion, density and momentum distributions, or tunneling rate, as well as explore mixed-dimensional systems~\cite{Nishida2008ufg, Nishida2009cie, Caracanhas2017fbm}, in which one component experiences a different dimensionality with respect to the other. Although progress has been already done in this context \cite{Lamporesi2010sim,Schaefer2018}, a wealth of predicted few- and many-body regimes still awaits experimental investigation.

As highlighted throughout this Review, the presence of a mass asymmetry between the mixture constituents significantly enriches the physical scenarios that can be accessed, enabling the observation of novel states of quantum matter in the context of few-body systems, impurity physics, and many-body physics. In this regard, the recent progress in producing novel FF mixtures \cite{Ravensbergen2018ado,Ciamei2022A} and FF Feshbach dimers \cite{Soave2023otf, Finelli2024ulc} is extremely appealing from several viewpoints. The intrinsic mismatch in the dispersion relations of the two components arising from the unequal masses could foster paradigmatic states of exotic superfluidity~\cite{Baarsma2010pam, Wang2017eeo, Pini2021bmf}. Moreover, stable few-body cluster states could be observed in both $^6$Li-$^{53}$Cr and $^{40}$K-$^{161}$Dy systems \cite{Kartavtsev2007let,Bazak2017,Pricoupenko2010,Liu2022A}. Besides representing an important step forward in the few-body context, these could represent a qualitative new opportunity to introduce non-perturbative elastic few-body correlations in fermionic media. In turn, these could promote novel types of impurity problems \cite{Mathy2011tma,Liu2022}, and trigger the emergence of exotic superfluid and normal states \cite{Endo2016,Liu2023}.

Another rapidly developing field, strongly connected with the one of quantum mixtures, concerns ultracold ground-state polar molecules \cite{Carr2009cau, Quéméner2012,Langen2023quantum},
which can be realized through a two-step process, based on the association of heteronuclear atom pairs into Feshbach dimers, which are subsequently transferred to deeply-bound states via coherent optical schemes. Such molecular gases, with which both Fermi and Bose quantum degeneracy has been already achieved \cite{Valtolina2020,Duda2023,Bigagli2024oob}, set the basis for countless applications and fundamental research: From novel quantum simulation and computation schemes based on the exploitation of long-ranged, tunable dipolar interactions, to controlled quantum chemistry and precision measurements.
 
To conclude, the field of ultracold quantum mixtures, established shortly after the advent of single-species quantum gases, has experienced impressive progress over the past two decades, evolving over the years into a variety of diverse, cross-disciplinary research directions, only partially covered by this work. Far from being exhausted, this research area still offers a wealth of possibilities for future investigation and further developments, and we hope that this short overview will also help motivate young scientists to pursue their experimental and theoretical studies on ultracold quantum mixtures.

\section*{Glossary}
\textbf{BCS-BEC crossover}\\
In fermionic mixtures, a crossover from a condensate of spatially large pairs coupled in momentum space, described by the Bardeen-Cooper-Schrieffer theory, to a condensate of tightly bound bosonic molecules by sweeping the interparticle interaction from attractive to repulsive, via, for example, a magnetic Feshbach resonance.\\ 
\textbf{Efimov states}\\
Few-body clusters of $N\geq$3 quantum particles supported by a long-range, scale-invariant $N$-body attractive potential that  originates from the combined effect of short-range resonant pairwise interactions and quantum statistics.\\
\textbf{Feshbach resonance}\\
In this context, a powerful tool used to tune the interactions between two atoms by applying an external, controllable field, typically a magnetic one. See Supplemental Information for more details.\\
\textbf{Miscibility}\\
Tendency of a mixture of two fluids to spatially overlap instead of separating into isolated domains. A mixture is miscible when the intercomponent repulsion is weaker than the average intracomponent one, and it is immiscible otherwise.\\
\textbf{Kartavtsev-Malykh trimers}\\
Three-body clusters, made of two identical heavy fermions and one light distinguishable particle, that arise in the presence of a resonant short-range pairwise interaction and a proper heavy-light mass ratio. Unlike Efimov states that exhibit a discrete scaling invariance, Kartavtsev-Malykh trimers feature a continuous scaling invariance and universal properties.\\
\textbf{Polaron}\\
State formed by a single impurity dressed by excitations of the medium it is embedded in: particle-hole or Bogoliubov for fermionic or bosonic media, respectively.\\
\textbf{Quantized vortex}\\
This topological structure is the characteristic signature of superfluidity. It is characterized by a localized density depletion around a line, where the wavefunction vanishes (vortex core). Around the vortex core the phase of the wavefunction accumulates a factor of $2\pi$. Correspondingly the atoms flow around the vortex core, with a velocity decreasing with the distance from the core.\\
\textbf{Soliton}\\
Localized nonlinear wave that tends to preserve its shape in time without spreading or shrinking, due to a balance between dispersion and nonlinearity of the medium.\\
\textbf{Spinor}\\
Multicomponent complex vector describing the wavefunction of a system with a nonzero spin. The dimension of the spinor is $2F+1$ for a system with spin $F$.

\section*{Acknowledgements}
We thank the members of the LENS Quantum Gases group in Florence and of the Pitaevskii BEC Center in Trento for fruitful discussions.
We thank R. Grimm for his scientific support and X. Cui, D.S. Petrov and P. Massignan for their critical reading of the manuscript and insightful discussions.
We also acknowledge financial support by
the PE0000023-NQSTI project by the Italian Ministry of University and Research, co-funded by the European Union — NextGeneration EU.
C.B and G.L. acknowledge financial support from Provincia Autonoma di Trento, and M.Z. acknowledges  financial support from the ‘Integrated infrastructure initiative in Photonic and Quantum Sciences’ I-PHOQS (CUP B53C22001750006).

\section*{Author contributions}
All authors equally contributed to all aspects of the article. 

\section*{Competing interests}
The authors declare no competing interests. 

\section*{Publisher’s note}
Springer Nature remains neutral with regard to jurisdictional claims in published maps and institutional affiliations.

\clearpage

\begin{center}
\textbf{\large Supplementary information: Quantum mixtures of ultracold gases of neutral atoms}
\end{center}

\setcounter{equation}{0}
\setcounter{figure}{0}
\setcounter{table}{0}

\section{Quantum mixtures}

Here below we report a list of different quantum mixtures mixtures of ultracold atoms which have been or are presently experimentally investigated. Table~\ref{tab:HomoMix}  collects systems with spin mixtures of a given atomic species (the internal state is labelled by the hyperfine quantum number $F$ and its projection $m_F$). Table~\ref{tab:HeteroMix}, instead, shows a collection of mixtures of different atoms and their mass imbalance is reported. 

Given the high number of experimental works published on quantum mixtures, the lists are clearly not exhaustive. 
We chose to cite one relevant paper per mixture per experimental group, so that the reader is aware of who/where such a mixture was or is studied and use it as a starting point for further and deeper search.

\begin{table}[h]
      \textbf{HOMONUCLEAR \ SPIN \ MIXTURES}\\
      \centering
      \vspace{3mm}
    \begin{tabular}{|c|c|c|c|c|}
         \hline
            ATOM & SPIN STATES $|F,m_F\rangle$  & STAT. & TOPICS & Refs. \\
             \hline
              &   &    &    & \\
               & \textbf{Two components}  &    &    & \\
               \hline
\cellcolor{red!10} $^4$He  & $|1,0\rangle$, $|1,+1\rangle$    & BB    & Impurity.   & \cite{Cayla2023oo1}\\
\hline
\cellcolor{red!15} $^6$Li  & $|1/2,\pm 1/2\rangle$    & FF    & BCS-BEC crossover, polarons,   & \cite{Jochim2003bec,Zwierlein2003oob, Bourdel2004eso,Partridge2006pap,Weimer2015cvi,Boll2016sad,Parsons2016srm,Cheuk2016oos,Kwon2021sea,Kuhn2020hfs,Cai2022pci,Helson2023dwo,Koch2023aqe,Rajkov2021fwi}\\
\cellcolor{red!15} $^6$Li  & $|\sfrac{1}{2},\sfrac{-1}{2}\rangle$, $|\sfrac{3}{2},\sfrac{-3}{2}\rangle$    & FF    & vortices, sound modes, transport,  & \cite{Holten2021ooc,Li2024oaq,Fabritius2024iet}\\
&&& (anti)ferromagnetism. &\\
\hline
\cellcolor{red!15} $^7$Li  & $|1,0\rangle$, $|1,+1\rangle$    & BB    & Spin transport.   & \cite{Jepsen2020spi}\\
\hline
\cellcolor{red!25} $^{23}$Na  & $|1,\pm 1\rangle$    & BB    & Sound modes, spin excitations, solitons.   & \cite{Kim2020oot,Cominotti2022oom,Chai2020msi}\\
\cellcolor{red!25} $^{23}$Na  & $|1,-1\rangle$, $|2,-2\rangle$    & BB    & Quantum simulation, ferromagnetism.   & \cite{Cominotti2023fia}\\
 \hline
\cellcolor{red!35} $^{39}$K  & $|1,-1\rangle$, $|1,0\rangle$    & BB    & LHY fluids, quantum droplets, Rabi coupling.   & \cite{Cabrera2017qld,Semeghini2018sbq,Skov2021ooa,Hammond2022ttb}\\
\cellcolor{red!35} $^{39}$K  & $|1,0\rangle$, $|1,+1\rangle$    & BB    & Polarons, quantum simulation, gauge theories.   & \cite{Jorgensen2016ooa,Frolian2022ra1}\\
 \hline
\cellcolor{red!35} $^{40}$K  & $|\sfrac{9}{2},\sfrac{-7}{2}\rangle$, $|\sfrac{9}{2},\sfrac{-9}{2}\rangle$    & FF    & BCS-BEC crossover, polarons, spin transport.   & \cite{Greiner2003eoa,Bardon2014tdd,Ness2020ooa}\\
 \hline
\cellcolor{red!45} $^{87}$Rb  & $|1,-1\rangle$, $|1,0\rangle$    & BB    & Impurity, transport, spin-orbit coupling.  & \cite{Palzer2009qtt,Lin2011soc}\\
\cellcolor{red!45} $^{87}$Rb  & $|1,-1\rangle$, $|2,-1\rangle$    & BB    & Artificial gauge fields.  & \cite{Aidelsburger2013roh}\\
\cellcolor{red!45} $^{87}$Rb  & $|1,-1\rangle$, $|2,+1\rangle$    & BB    & Vortices, miscibility, entanglement, spin squeezing.  & \cite{Matthews1999via,Nicklas2015nds,Colciaghi2023epr,Huang2023oos}\\
\cellcolor{red!45}  $^{87}$Rb  & $|1,0\rangle$, $|1,+1\rangle$    & BB    & Solitons, superfluidity.  & \cite{Hamner2011god,Moulder2012qsd}\\
\cellcolor{red!45} $^{87}$Rb  & $|1,0\rangle$, $|2,0\rangle$    & BB    & Solitons.  & \cite{BakkaliHassani2021roa}\\
\cellcolor{red!45} $^{87}$Rb  & $|1,+1\rangle$, $|1,-1\rangle$    & BB    & Spin Mott insulator.  & \cite{deHond2022pot}\\
\cellcolor{red!45} $^{87}$Rb  & $|1,+1\rangle$, $|2,-1\rangle$    & BB    & Vortices, miscibility.  & \cite{Nicklas2011rfi}\\
 \hline
\cellcolor{red!60} $^{162}$Dy  & $|8,-8\rangle$, $|8,-7\rangle$    & BB    &  Collisional properties.  & \cite{Lecomte2024pas}\\
 \hline
\cellcolor{red!65} $^{167}$Er  & $|\sfrac{19}{2},\sfrac{-19}{2}\rangle$, $|\sfrac{19}{2},\sfrac{-17}{2}\rangle$    & FF    &  Collisional properties, ultracold mixture.  & \cite{Baier2018roa}\\
    \hline
 &   &    &    & \\
 & \textbf{Multicomponent}  &    &    & \\
     \hline
\cellcolor{red!15} $^{6}$Li  & $ |\sfrac{1}{2}, \sfrac{\pm 1}{2}\rangle $, $|\sfrac{3}{2},\sfrac{-3}{2}\rangle$     & 3 F    &  Few-body physics.  & \cite{Wenz2009uti,Huckans2009tbr,Schumacher2023ooa}\\
\hline
\cellcolor{red!25}   $^{23}$Na  & Full spinor $F=1$     & 3 B    &  Spinor miscibility, vortices.  & \cite{StamperKurn1998oco,Seo2015hqv,Chai2020msi,JimenezGarcia2019sfa}\\
 \hline
\cellcolor{red!35} $^{52}$Cr  &  $|3, m_F<0 \rangle$  & 3 B    & Spin dynamics.   & \cite{Naylor2016cbb}\\
 \hline
\cellcolor{red!45}  $^{87}$Rb  & Full spinor $F=1$     & 3 B    &  Spinor miscibility, vortices, solitons, spin dynamics.  & \cite{Barrett2001aof,Sadler2006ssb,Lannig2020cot,Meyer-Hoppe2023esp}\\
\cellcolor{red!45} $^{87}$Rb  & Full spinor $F=2$     & 5 B    &  Spinor dynamics.  & \cite{Schmaljohann2004dof}\\
 \hline
\cellcolor{red!50}  $^{87}$Sr  & Full spinor $F=9/2$   & 10 F    & SU(N)-symmetric Fermi gases.  & \cite{DeSalvo2010dfg,Stellmer2013poq,Sonderhouse2020toa}\\
 \hline
\cellcolor{red!60}  $^{162}$Dy  & Full spinor $F=8$     & 17 B    &  Synthetic dimensions.  & \cite{Bouhiron2024roa}\\
\hline
\cellcolor{red!65} $^{173}$Yb  & Full spinor $F=3/2$     & 6 F    &  SU(N)-symmetric Fermi gases.  & \cite{Pagano2014aod,Taie2010roa}\\
\hline
    \end{tabular}
    \vskip -15pt
    \caption{Survey of homonuclear spin mixture experiments. The table reports the atomic species and isotope of each spin mixture from the lightest (light red) to the heaviest (dark red), the involved internal spin states expressed in terms of the total angular momentum $F$ and its projection $m_F$, the resulting statistics of the mixture, some of the topics studied, and references from different groups. In the upper part of the table, two-component mixtures are considered, whereas mixtures with more than two components are reported in the lower part.}
    \label{tab:HomoMix}
\end{table}

\begin{table}[hbt!] 
 \textbf{HETERONUCLEAR \ MIXTURES}\\
    \centering   
    \vspace{3mm}
    \begin{tabular}{|c|c|c|c|c|c|}
       \hline
            ATOM 1 & ATOM 2 & MASS IMB. & STAT. & TOPICS & Refs. \\
             \hline
\cellcolor{red!10} $^3$He     & \cellcolor{red!10} $^4$He    & 1.3   & FB    &    Ultracold mixture.  & \cite{McNamara2006dbf} \\
             \hline
\cellcolor{red!10} $^4$He     & \cellcolor{red!45} $^{87}$Rb    & 21.7   & FB    &   Ultracold mixture.   & \cite{Flores2017auo} \\
             \hline
\cellcolor{red!15}$^6$Li     & \cellcolor{red!15} $^7$Li    & 1.2   & FB    &   Double degeneracy, superfluidity, hydrodynamics.   & \cite{Ferrierbarbut2014amo,Truscott2001oof} \\
\cellcolor{red!15} $^6$Li     & 
\cellcolor{red!25} $^{23}$Na    & 3.8   & FB    &   Double degeneracy, impurity physics, molecules.   & \cite{Hadzibabic2002tsm,Scelle2014mco,Rvachov2017llu} \\
\cellcolor{red!15} $^6$Li     & \cellcolor{red!35} $^{40}$K    & 6.6   & FF    &   Molecules, impurity physics.   & \cite{Yang2020spt,Kohstall2012mac} \\
\cellcolor{red!15} $^6$Li     & \cellcolor{red!35} $^{41}$K    & 6.8   & FB    &   Vortices, impurity physics. & \cite{Yao2016ooc,Baroni2023mib} \\
\cellcolor{red!15} $^6$Li     & \cellcolor{red!40} $^{53}$Cr    & 8.8   & FF    &   Molecules. & \cite{Finelli2024ulc} \\
\cellcolor{red!15} $^6$Li     & 
\cellcolor{red!45} $^{84}$Sr    & 14.0   & FB    &   Double degeneracy. & \cite{Ye2020ddb} \\
\cellcolor{red!15} $^6$Li     & \cellcolor{red!45} $^{87}$Rb    & 14.4   & FB    &   Double degeneracy. & \cite{Silber2005qdm} \\
\cellcolor{red!15} $^6$Li     & \cellcolor{red!55} $^{133}$Cs    & 22.1   & FB    &   Few-body physics. & \cite{Repp2013ofi,Tung2014gso} \\
\cellcolor{red!15} $^6$Li     & \cellcolor{red!55} $^{138}$Ba$^+$    & 22.9   & F+    &   Collisional properties.  & \cite{Weckesser2021oof} \\
\cellcolor{red!15} $^6$Li     & \cellcolor{red!60} $^{166}$Er    & 27.6   & FB    &   Collisional properties.  & \cite{Schaefer2022fro} \\
\cellcolor{red!15} $^6$Li     & \cellcolor{red!60} $^{168}$Er    & 27.9   & FB    &   Collisional properties.  & \cite{Schaefer2022fro} \\
\cellcolor{red!15} $^6$Li     & \cellcolor{red!65} $^{171}$Yb$^+$    & 28.4   & F+    &   Collisional properties.  & \cite{Feldker2020bgc} \\
\cellcolor{red!15} $^6$Li     & \cellcolor{red!65} $^{174}$Yb    & 28.9   & FB    &   Ultracold mixture, few-body physics. & \cite{Okano2010smo,Green2020fri} \\
 \hline
\cellcolor{red!15} $^7$Li     & \cellcolor{red!45} $^{84}$Sr    & 12.0   & BB    &   Double degeneracy & \cite{Ye2020ddb} \\
\cellcolor{red!15} $^7$Li     & \cellcolor{red!45} $^{87}$Rb    & 12.4   & BB    &   Collisional properties, few-body physics. & \cite{Maier2015era,Fang2020cdr} \\
\cellcolor{red!15} $^7$Li     & \cellcolor{red!55}  $^{133}$Cs    & 18.9   & BB    &   Ultracold mixture, few-body physics. & \cite{Mosk2001mou,Chen2023dsb} \\
\cellcolor{red!15} $^7$Li     & \cellcolor{red!60} $^{166}$Er    & 23.7   & BB    &   Collisional properties. & \cite{Schaefer2022fro} \\
\cellcolor{red!15} $^7$Li     & \cellcolor{red!60} $^{168}$Er    & 23.9   & BB    &   Collisional properties. & \cite{Schaefer2022fro} \\
\hline
\cellcolor{red!25} $^{23}$Na     & \cellcolor{red!35} $^{39}$K    & 1.7   & BB    &   Molecules. & \cite{Voges2020ugo} \\
\cellcolor{red!25} $^{23}$Na     & \cellcolor{red!35} $^{40}$K    & 1.7   & BF    &   Molecules, impurity physics. & \cite{Park2015udg,Yang2022coa,Chen2024ufl,Duda2023tfa} \\
\cellcolor{red!25}  $^{23}$Na     & 
\cellcolor{red!35} $^{41}$K    & 1.8   & BB    &   Double degeneracy. & \cite{Chang2024dsb} \\
\cellcolor{red!25}  $^{23}$Na     & \cellcolor{red!45} $^{87}$Rb    & 3.8   & BB    &   Collisional properties, molecules. & \cite{Wang2013oof} \\
\cellcolor{red!25}  $^{23}$Na     & \cellcolor{red!55} $^{133}$Cs    & 5.8   & BB    &   Molecules, molecular BEC. & \cite{Liu2018fso,Bigagli2024oob} \\
\hline
\cellcolor{red!35} $^{39}$K     & \cellcolor{red!45} $^{87}$Rb    & 2.2   & BB    &   Collisional properties, few-body physics, double degeneracy. & \cite{Simoni2008ntm,Wacker2016utb,Mi2021pod} \\
\cellcolor{red!35} $^{39}$K     & \cellcolor{red!45}  $^{133}$Cs    & 3.4   & BB    &   Collisional properties. & \cite{Grobner2017ooi} \\
\hline
\cellcolor{red!35} $^{40}$K     & \cellcolor{red!45} $^{87}$Rb    & 2.2   & FB    &   Double degeneracy, collisional properties, molecules. & \cite{Roati2002fbq,Ospelkaus2006idd,Ni2008ahp,Thomas2018oob} \\
\cellcolor{red!35} $^{40}$K     & \cellcolor{red!60} $^{161}$Dy    & 4.0   & FF    &   Molecules. & \cite{Soave2023otf} \\
\hline
\cellcolor{red!35} $^{41}$K     & \cellcolor{red!45} $^{87}$Rb    & 2.1   & BB    &  Double degeneracy, quantum droplets, few-body physics.  & \cite{Modugno2002tas,DErrico2019ooq,Wacker2016utb,Elliott2023qgm} \\
\hline
\cellcolor{red!45} $^{84}$Sr   &  \cellcolor{red!45} $^{87}$Rb     & $\simeq$1   & BB    &    Double degeneracy. & \cite{Pasquiou2013qdm} \\
\cellcolor{red!45} $^{84}$Sr     & \cellcolor{red!45} $^{87}$Sr    & $\simeq$1  & BF    &  Double degeneracy.  & \cite{Tey2010ddb} \\
\hline
\cellcolor{red!45} $^{85}$Rb     & \cellcolor{red!45} $^{87}$Rb    & $\simeq$1   & BB    &   Double degeneracy, collisional properties. & \cite{Papp2008tmi,Cui2017oob} \\
\cellcolor{red!45} $^{85}$Rb     & \cellcolor{red!55}$^{133}$Cs    & 1.6   & BB    &   Collisional properties. & \cite{Cho2013fso} \\
\hline
\cellcolor{red!45} $^{86}$Sr     & \cellcolor{red!45} $^{88}$Sr    & $\simeq$1  & BB    &  Ultracold mixture.  & \cite{Poli2005cat} \\
\hline
\cellcolor{red!45} $^{87}$Rb     & \cellcolor{red!45} $^{87}$Rb$^+$    & $\simeq$1   & B+    &  Collisional properties, ion transport. & \cite{Dieterle2021toa} \\
\cellcolor{red!45} $^{87}$Rb     & \cellcolor{red!45} $^{87}$Sr    & $\simeq$1   & BF    &   Collisional properties. & \cite{Barbe2018oof} \\
\cellcolor{red!45} $^{87}$Rb     & \cellcolor{red!45} $^{88}$Sr    & $\simeq$1   & BB    &   Collisional properties, double degeneracy. & \cite{Pasquiou2013qdm,Barbe2018oof} \\
\cellcolor{red!45} $^{87}$Rb     & \cellcolor{red!55} $^{133}$Cs    & 1.5   & BB    &   Collisional properties, molecules. & \cite{Takekoshi2014uds,Molony2014cou,Schmidt2019tsa} \\
\cellcolor{red!45} $^{87}$Rb     & \cellcolor{red!65} $^{174}$Yb$^+$    &  2.0  & B+    &   Collisional properties. & \cite{Zipkes2010ats} \\
\hline
\cellcolor{red!55} $^{133}$Cs     & \cellcolor{red!65} $^{174}$Yb   &  1.3  & BB    &  Double degeneracy, collective dynamics. & \cite{Wilson2021doa} \\
\hline
\cellcolor{red!60} $^{161}$Dy     & \cellcolor{red!60} $^{162}$Dy   &  $\simeq$1  & FB    &  Double degeneracy. & \cite{Lu2012qdd} \\
\cellcolor{red!60} $^{161}$Dy     & \cellcolor{red!60} $^{166}$Er   &  $\simeq$1  & FB    &  Ultracold mixture. & \cite{Trautmann2018dqm} \\
\cellcolor{red!60} $^{161}$Dy     & \cellcolor{red!63} $^{168}$Er   &  $\simeq$1  & FB    &  Double degeneracy. & \cite{Trautmann2018dqm} \\
\cellcolor{red!60} $^{161}$Dy     & \cellcolor{red!63} $^{170}$Er   &  $\simeq$1  & FB    &   Ultracold mixture. & \cite{Trautmann2018dqm} \\
\hline
\cellcolor{red!60}$^{162}$Dy     & \cellcolor{red!63} $^{166}$Er   &  $\simeq$1  & BB    &  Ultracold mixture. & \cite{Trautmann2018dqm} \\
\cellcolor{red!60}$^{162}$Dy     & \cellcolor{red!63} $^{168}$Er   &  $\simeq$1  & BB    &  Double degeneracy. & \cite{Trautmann2018dqm} \\
\cellcolor{red!60} $^{162}$Dy     & \cellcolor{red!63} $^{170}$Er   &  $\simeq$1  & BB    &  Double degeneracy. & \cite{Trautmann2018dqm} \\
\hline
\cellcolor{red!60}$^{164}$Dy     & \cellcolor{red!63} $^{166}$Er   &  $\simeq$1  & BB    &  Double degeneracy. & \cite{Trautmann2018dqm} \\
\cellcolor{red!60} $^{164}$Dy     & \cellcolor{red!63} $^{168}$Er   &  $\simeq$1  & BB    &  Double degeneracy. & \cite{Trautmann2018dqm} \\
\cellcolor{red!60} $^{164}$Dy     &  \cellcolor{red!63}$^{170}$Er   &  $\simeq$1  & BB    &  Double degeneracy. & \cite{Trautmann2018dqm} \\
\hline
\cellcolor{red!62} $^{168}$Er  & \cellcolor{red!65} $^{174}$Yb &  $\simeq$1      & BB    &  Double degeneracy.  & \cite{Schafer2023roa}\\
\hline
\cellcolor{red!65} $^{171}$Yb  & \cellcolor{red!65} $^{173}$Yb &  $\simeq$1      & FF    &  Double degeneracy, SU(N) symmetry.  & \cite{Taie2010roa}\\
\cellcolor{red!65} $^{171}$Yb  & \cellcolor{red!65} $^{172}$Yb$^+$ &  $\simeq$1      & F+    &  Collisional properties.  & \cite{Grier2009ooc}\\
\hline
\cellcolor{red!65} $^{172}$Yb  & \cellcolor{red!65} $^{174}$Yb$^+$ &  $\simeq$1      & B+    &  Collisional properties.  & \cite{Grier2009ooc}\\
\hline
\cellcolor{red!65} $^{174}$Yb  & \cellcolor{red!65} $^{172}$Yb$^+$ &  $\simeq$1      & B+    &  Collisional properties.  & \cite{Grier2009ooc}\\
\hline
    \end{tabular}
    \vskip -15pt
        \caption{Survey of heteronuclear mixture experiments. The table shows the two atomic species and isotopes involved in the mixture from lighter (light red) to heavier (dark red), its mass imbalance, the statistical mix, some studied topics, and relevant representative references.}
    \label{tab:HeteroMix}
\end{table}

\clearpage

\section{Feshbach resonances}

\textit{Feshbach resonances} (FRs) are an inestimable tool that allows control on the scattering length, and thus on the interactions, between particles. While we refer the reader to Refs.~\cite{Chin2010fri, Kokkelmans2014fri} for a detailed explanation of FRs in ultracold gases, here we briefly review the main concepts of $s$-wave FRs, mostly following Ref.~\cite{Chin2010fri}. 

\begin{figure}[h!!]
\begin{centering}
    \includegraphics[width = 1\textwidth]{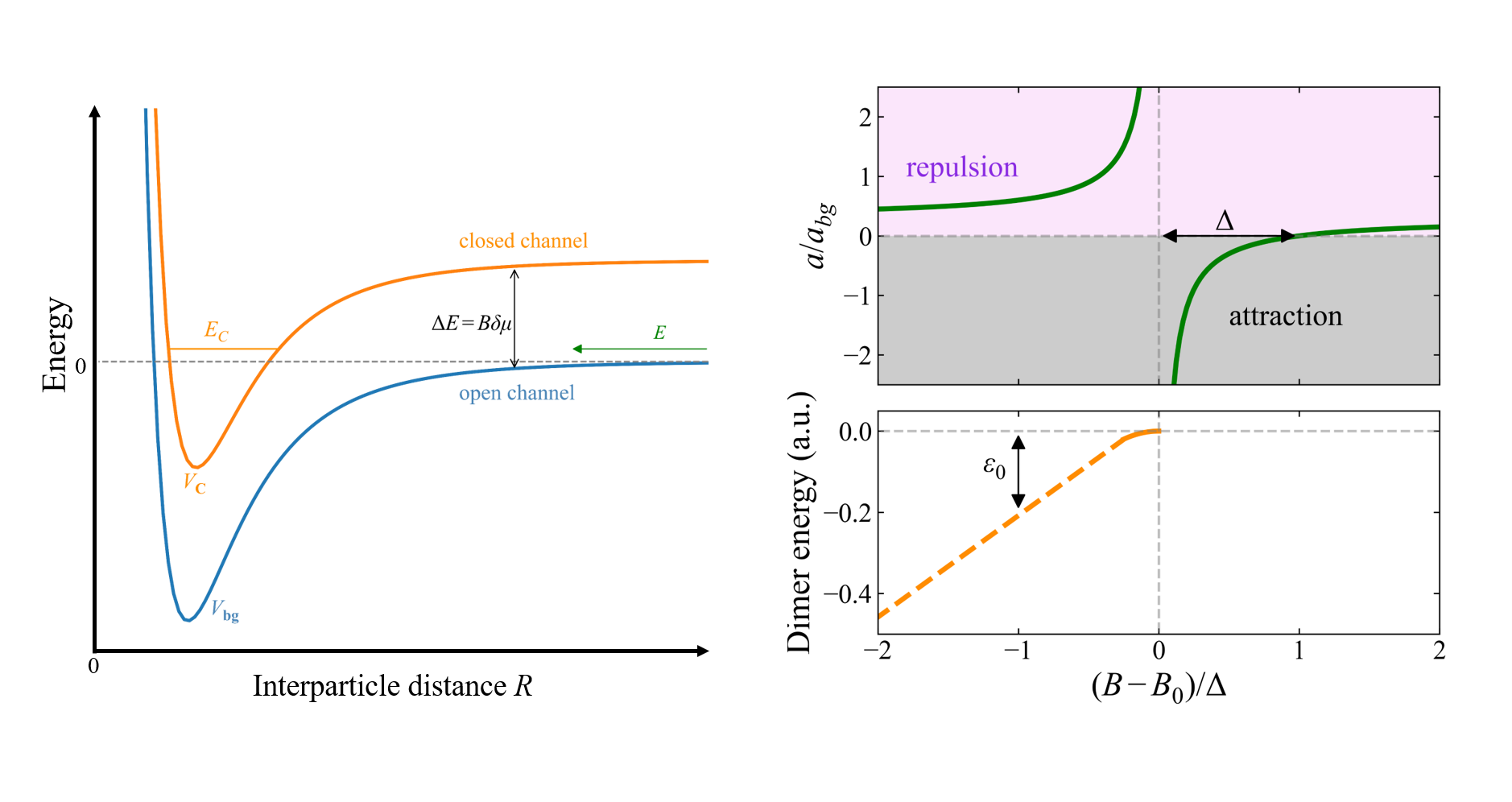}
\caption[Sketch of a Feshbach resonance]{\label{fig:channels_model} \textbf{Sketch of a Feshbach resonance.} \textbf{Left:} Two-channel model for a Feshbach resonance. When two atoms collide at energy $E$ in the open channel they resonantly couple to a molecular bound state of energy $E_c$ in the closed channel. In magnetic FR the energy difference $\Delta E$ between the two channels can be tuned thanks to an  external applied magnetic field $B$. \textbf{Right:} scattering length $a$ (upper panel) and molecular state energy (lower panel) $vs$ the applied external magnetic field in the vicinity of a FR.}
\end{centering}
\end{figure}

The basic idea can be pictured with a two-channel model for the scattering process between two particles, and is depicted in Fig.~\ref{fig:channels_model}. We consider two molecular potentials, $V_{\text{bg}}(R)$ and $V_{\text{C}}(R)$, as functions of the interparticle distance $R$. $V_{\text{bg}}(R)$ asymptotically connects to the two free particles and, for small collisional energy $E$, represents the open (or entrance) channel. $V_{\text{C}}(R)$ represents the energetically forbidden closed channel, and can support bound molecular states with energy $E_{\text{C}}(R)$ close to the threshold of the open channel. 
If the bound state in the closed channel energetically approaches the scattering state in the open channel, the colliding atoms resonantly couple to this bound state and their scattering length diverges. This is the mechanism behind the occurrence of FR. If there is no coupling between the two channels, the existence of the bound state in the closed channel has no effect on the scattering in
the open channel. On the other hand, in the presence of small
coupling, the scattering length will be large and positive if the bound state in the closed channel is just
below the threshold of the continuum spectrum in the
open channel, and large and negative if it is above such threshold. Note that a weak coupling can be enough to lead to a strong mixing between the two channels.
If the two channels have different magnetic moments ($\delta \mu \neq 0$), the energy difference between them can be tuned by applying an external magnetic field $B$, according to $\Delta E = B\delta\mu$. This tunability is what makes FR particularly appealing: Simply applying an external field we are able to tune the interactions between two atoms at will. 
Ref.~\cite{Moerdijk1995riu} introduced a simple dependence between the scattering length of the two colliding particles and the applied magnetic field:
\begin{align}
  a = a_\text{bg}\left(1-\frac{\Delta}{B-B_0}\right),
\end{align}
where $a_\text{bg}$ is the off-resonant value of the scattering length associated with $V_{\text{bg}}(R)$, $B_0$ the center of the resonance, and $\Delta$ its width. Note that for $B=B_0$ the scattering length diverges and we reach the unitarity limit.

When the two channel resonantly mix, an avoid crossing between the scattering and the molecular states occurs, giving rise to two separated branches: A repulsive branch in which the scattering atoms repel each other, and an energetically lower attractive branch, in which the atoms are attracted between each other. This interaction grows with approaching the center $B_0$ of the FR, where the scattering length diverges and a \textit{dressed} molecular state occurs. For large positive values of the scattering length, the binding energy of this dressed molecular state is given by
\begin{equation}
    \varepsilon_0 = -\frac{\hslash r^2}{2 m_r a^2}
\end{equation}
and depends quadratically on the magnetic detuning $B-B_0$, because of the strong influence of the mixing of the channels.
Away from the resonance center, in the region where $a>0$, the energy of the weak bound molecular state varies linearly with the applied magnetic field with a slope given by $\delta \mu$ (see Fig.~S~\ref{fig:channels_model}). \\

Note that, while typically magnetic Feshbach resonances are used, which are based on Zeeman shifting a bound molecular state into resonance with the scattering state, other schemes are possible.
A powerful example are optical Feshbach resonances \cite{Fedichev1996ion}, which allow to tune the interactions thanks to the presence of electronically excited states. They were observed in alkali atoms ($^{23}$Na and $^{87}$Rb) \cite{Fatemi2000ooo, Theis2004tts, Thalhammer2005iao}, alkali-earth atoms ($^{88}$Sr) \cite{Zelevinsky2006nlp, Blatt2011moo}, and non-magnetic Ln atoms ($^{172}$Yb, $^{176}$Yb) \cite{Enomoto2008ofr}, reaching submicron spatial control of interatomic interactions in a BEC of $^{174}$Yb \cite{Yamazaki2010ssm}. Involving laser light, instead of magnetic fields, these resonances are addressed much faster than magnetic ones, avoiding sweeping and fluctuations of currents in the coils. One trick to fast address magnetic FR consists in exploiting state-dependent light shift to quickly shift the magnetic resonance, as demonstrated in \cite{Bauer2009coa, Jagannathan2016oco, Cetina2016umb}.
Also radio-frequency (RF) \cite{Tscherbul2010rfi, Hanna2010cam,Owens2016cfr, Ding2017eco} and microwaves \cite{Papoular2010mif} FR are possible, and the RF radiation can be use for dressing multiple Feshbach resonances \cite{Kaufman2009rfd}. 
Moreover, orbital interaction-induced Feshbach resonances were observed in $^{173}$Yb~\cite{Hoefer2015ooa} and are based on the Zeeman shift of different nuclear spin states of the atoms.\\
Finally, resonant scattering can be achieved also with the manipulation of the confinement of the atomic cloud (confinement-induced resonances) \cite{Olshanii1998asi}, as it has been shown in 1D \cite{Guenter2005pwi, Haller2010cir} ($^{40}$K and $^{133}$Cs, respectively), in 2D \cite{Froehlich2011rfs} ($^{40}$K), and recently in 0D \cite{Lee2023spd, Capecchi2023ofc} ($^7$Li and $^{133}$Cs, respectively). 


\end{document}